\documentclass[pre,twocolumn,aps,floatfix,superscriptaddress,10pt]{revtex4-2}
\usepackage{graphicx}
\usepackage{amssymb}
\usepackage{amsmath}
\usepackage[usenames]{color}
\usepackage[normalem]{ulem}
\DeclareMathOperator{\sech}{sech}

\usepackage{hyperref}
\hypersetup{colorlinks,
 linkcolor=blue,%
 citecolor=blue,%
 urlcolor=blue
}

\begin{document}

\title{Stability and dynamics of dark-bright solitons in spin-orbit- and Rabi-coupled binary Bose-Einstein condensates}
\author{K. Rajaswathi}
\affiliation{Department of Physics, Bharathidasan University, Tiruchirappalli 620024, Tamil Nadu, India}

\author{R. Ravisankar}
\affiliation{Department of Mechanics and Aerospace Engineering, Southern University of Science and Technology, Shenzhen 518055, China}

\author{R. Radha}
\affiliation{Centre for Nonlinear Science (CeNSc), Government College for Women (Autonomous), Kumbakonam 612001, Tamil Nadu, India}

\author{P. K. Mishra}
\affiliation{Department of Physics, Indian Institute of Technology Guwahati, Guwahati 781039, Assam, India}

\author{P. Muruganandam}
\affiliation{Department of Physics, Bharathidasan University, Tiruchirappalli 620024, Tamil Nadu, India}
\affiliation{Department of Medical Physics, Bharathidasan University, Tiruchirappalli 620024, Tamil Nadu, India}


\begin{abstract}
We investigate the stability and nonlinear dynamics of dark--bright solitons in a one-dimensional binary Bose--Einstein condensate subjected to synthetic spin--orbit and Rabi couplings. In the absence of spin--orbit coupling, we map the coupled Gross--Pitaevskii equations onto the integrable Manakov model to obtain exact dark--bright soliton solutions, providing a rigorous theoretical benchmark. We demonstrate that finite spin--orbit coupling breaks integrability by inducing spin-dependent phase gradients that drive component-wise spatial separation and intrinsic density oscillations. By contrast, coherent Rabi driving enforces phase locking between spin components and supports robust breather-like excitations. Furthermore, we derive analytical continuity relations for mass and spin current densities, mapping the internal spin dynamics onto an internal Josephson-junction framework in which the gauge field acts as a continuous spatial momentum bias. Using imaginary-time propagation together with Bogoliubov--de Gennes analysis, we systematically characterise ground-state phases and excitation spectra for both symmetric and asymmetric interaction regimes in homogeneous and harmonically trapped systems. Real-time simulations further demonstrate that synthetic gauge fields and interaction quenches drive the system far from equilibrium, triggering modulational-instability-induced multi-soliton fragmentation, breathing stripe patterns, and non-equilibrium transport. Our results highlight the interplay of synthetic gauge fields, external confinement, and interaction engineering as powerful tools for controlling the stability and internal dynamics of multicomponent quantum fluids.
\end{abstract}

\maketitle

\section{Introduction}

Over the past few decades, Bose--Einstein condensates (BECs) have emerged as a versatile platform for exploring a broad range of phenomena in quantum and condensed matter physics~\cite{Pethick2008, StamperKurn2013, Pitaevskii2010}. Their high degree of experimental controllability enables systematic investigation of non-linear excitations, coherence, superfluidity, and collective dynamics under highly tunable conditions. The static and dynamical properties of these dilute ultracold gases are well captured within the mean-field framework of the Gross--Pitaevskii (GP) equation, which has been proven to be remarkably successful in describing both equilibrium ground states and non-equilibrium wave propagation~\cite{Leggett2001, Kawaguchi2012, Pitaevskii2016}.

A pivotal development in this context is the experimental realisation of synthetic spin--orbit (SO) coupling in ultracold systems~\cite{Lin2011, Galitski2013}. Achieved via Raman transitions between internal hyperfine states, synthetic SO coupling generates momentum-dependent interactions between spin components, opening new avenues for investigating spin transport~\cite{Lin2011, Deng2024, Hasan2010}, topological phases~\cite{Bloch2012, Amin2016}, and non-linear matter-wave dynamics~\cite{Amin2016, Cui2019, Chen2020, Zhang2016, Yanchao2023, Sanz2022}. Crucially, the inclusion of SO coupling fundamentally modifies vector soliton dynamics by breaking Galilean invariance~\cite{Lin2011, Kevrekidis2007}. In multicomponent systems, such as dark--bright and dark--dark solitons, this symmetry breaking gives rise to novel dynamical phenomena, including shape transformations, internal state beating, and Zitterbewegung-like oscillations. Multiscale expansion techniques, variational methods, and numerical simulations have further elucidated how internal modes and Goldstone excitations govern these oscillatory states~\cite{Achilleos2014, Weizhu2013, Alotaibi2017}.

Beyond isolated soliton dynamics, the interplay of SO coupling, coherent Rabi driving, and non-linear atomic interactions enables a rich variety of wave-mixing processes and complex phases. For instance, spontaneous degenerate four-wave mixing in SO-coupled systems permits phase-matched wave generation governed by energy--momentum conservation, offering a controllable route for generating correlated atomic beams~\cite{Hung2020}. Furthermore, interaction quenches involving combined SO and Rabi couplings provide a mechanism to tune spin transport, trigger miscible--immiscible phase transitions, and generate plane-wave or stripe phases~\cite{Ravisankar2020_2, Ravisankar2025, Fei-Yan2024, Vinayagam2025}. Similar non-linear structures have also been explored in higher-spin condensates~\cite{He2023, Hui2021, Wang2024}, optical lattice geometries~\cite{Yang2023, Yanchao2023}, and systems with spatially modulated interactions mapped to integrable models~\cite{BeloboBelobo2023}.

Despite these significant advances, a comprehensive understanding of how soliton stability and non-linear dynamics can be systematically controlled through the combined effects of synthetic SO coupling, Rabi driving, and external harmonic confinement remains incomplete. In particular, the precise mechanisms governing the non-equilibrium evolution, internal breathing, and fragmentation of dark--bright solitons under interaction quenches in quasi-one-dimensional geometries warrant further detailed investigation.

In this work, we address these questions by presenting a systematic study of the stability and non-linear dynamics of vector dark--bright solitons in a pseudo-spin-$1/2$ Bose--Einstein condensate subjected to Raman-generated synthetic SO coupling and coherent Rabi driving. By combining exact analytical derivations in the integrable limit, Bogoliubov--de Gennes (BdG) stability analysis, and direct real-time GP numerical simulations, we elucidate how the interplay between synthetic gauge fields, inter-component interaction anisotropy, and external confinement shapes both stationary bound states and non-equilibrium dynamics.

Specifically, we establish exact dark--bright soliton solutions within the Manakov limit via spatial gauge transformations and explore how finite SO coupling breaks integrability, introducing spin-dependent phase gradients and structural deformities. Through imaginary-time propagation and BdG spectral analysis, we map out the linear stability regimes of these localised states. Real-time dynamical simulations further reveal a rich spectrum of non-linear phenomena, including centre-of-mass oscillations, internal breathing, stripe-phase modulations, component separation, and the formation of robust soliton trains under interaction quenches.

The remainder of this manuscript is organized as follows. In section \ref{sec:2}, we formulate the coupled quasi-1D Gross--Pitaevskii model. Section \ref{sec:3} details the mapping to the integrable Manakov hierarchy and derives the exact dark--bright soliton solutions along with their associated mass and spin currents. In section \ref{sec:4}, we present the BdG stability analysis and excitation spectra in harmonic traps. Section \ref{sec:5} discusses direct numerical simulations of the non-linear dynamics under interaction quenches and in free space, and section \ref{sec:6} summarises our primary conclusions.

\section{The mean-field model}
\label{sec:2}
The Gross--Pitaevskii (GP) equation provides an excellent framework for studying the ground states and their corresponding dynamical properties of Bose--Einstein condensates. In the case of a pseudo-spin-$1/2$ condensate subjected to synthetic spin–orbit and Rabi couplings~\cite{Ravisankar2020_2, Sarkar2025}, the system is described by two coupled GP equations governing the macroscopic wave functions $\psi_{\uparrow}(x,t)$ and $\psi_{\downarrow}(x,t)$, which represent the spin-up and spin-down components of the condensate, respectively. These equations take the form
\begin{subequations}
\label{eq:gpsoc:1}
\begin{align}
\mathrm{i}\partial_t \psi_\uparrow = & \left[ -\frac{1}{2}\partial_x^2 - \mathrm{i}k_L\partial_x + V(x) + g_{\uparrow\uparrow}|\psi_\uparrow|^2 + g_{\uparrow\downarrow}|\psi_\downarrow|^2 \right]\psi_\uparrow \notag \\ & \quad + \Omega \psi_\downarrow \label{eq:gpsoc:1a}  
\end{align}
\begin{align}
\mathrm{i} \partial_{t} \psi_{\downarrow} = & \left[ -\frac{1}{2}\partial_{x}^2 + \mathrm{i} k_L \partial_{x} + V(x) + g_{\downarrow \uparrow} \lvert \psi_{\uparrow} \rvert^2 + g_{\downarrow \downarrow} \lvert \psi_{\downarrow} \rvert^2 \right] \psi _{\downarrow} \notag \\ & \quad  + \Omega \psi_{\uparrow}, \label{eq:gpsoc:1b}
\end{align}
\end{subequations}%
Here, $V(x)$ denotes the external trapping potential, $k_{L}$ characterises the strength of the spin-orbit coupling, and $\Omega$ is the Rabi coupling strength. The coefficients $g_{\uparrow\uparrow}$ and $g_{\downarrow\downarrow}$ correspond to the intraspecies interaction strengths, while $g_{\uparrow\downarrow}$ describes the interspecies interaction between the two components.

The ground state is obtained using the imaginary-time propagation method. During this procedure, the SO and Rabi couplings are switched off ($k_L = 0$, $\Omega = 0$), and the normalisation of each component is fixed according to
\begin{align}
 N_\uparrow = \int \lvert \psi_\uparrow \rvert^2 dx = N_1, \quad \text{and} \quad 
 N_\downarrow = \int \lvert \psi_\downarrow \rvert^2 dx = N_2. \label{eq:norm}
\end{align}
This approach ensures convergence to a nontrivial stationary state with prescribed populations in each component. In contrast, during real-time evolution, the presence of Rabi coupling allows interconversion between spin states, so that only the total norm $N = N_\uparrow + N_\downarrow$ is conserved.

To compute the chemical potentials of the individual components computationally, we express the wave functions as, $\psi_{\uparrow, \downarrow}(x,t) = \mathrm{e}^{-\mathrm{i} \mu_{\uparrow, \downarrow} t} (\psi_{\uparrow, \downarrow R} + \mathrm{i}\psi_{\uparrow, \downarrow I})$. Taking the inner product with the complex conjugate and integrating, the chemical potentials at $t = 0$ are obtained as
\begin{subequations}
\label{eq:gpsoc2:mu}
\begin{align}
\mu_\uparrow =&\, \frac{1}{N_\uparrow} \int \Bigg\{ \frac{1}{2} \left\lvert \frac{\partial \psi_\uparrow}{\partial x} \right\rvert^2 + k_L \left(\psi_{\uparrow R} \frac{\partial \psi_{\uparrow I}}{\partial x} - \psi_{\uparrow I} \frac{\partial \psi_{\uparrow R}}{\partial x} \right) \notag \\
& \quad + \Big[ V(x) + g_{\uparrow \uparrow} \lvert \psi_{\uparrow} \rvert^2 + g_{\uparrow \downarrow} \lvert \psi_{\downarrow} \rvert^2 \Big] \lvert \psi_{\uparrow} \rvert^2 \notag \\
& \quad + \Omega \left(\psi_{\uparrow R} \psi_{\downarrow R} + \psi_{\uparrow I} \psi_{\downarrow I}\right) \Bigg\} dx, \\
\mu_\downarrow = & \, \frac{1}{N_\downarrow} \int \Bigg\{ \frac{1}{2} \left\lvert \frac{\partial \psi_\downarrow}{\partial x} \right\rvert^2 - k_L \left(\psi_{\downarrow R} \frac{\partial \psi_{\downarrow I}}{\partial x} - \psi_{\downarrow I} \frac{\partial \psi_{\downarrow R}}{\partial x} \right) \notag \\
& \quad + \Big[ V(x) + g_{\downarrow \uparrow} \lvert \psi_{\uparrow} \rvert^2 + g_{\downarrow \downarrow} \lvert \psi_{\downarrow} \rvert^2 \Big] \lvert \psi_{\downarrow} \rvert^2 \notag \\
& \quad  + \Omega \left(\psi_{\uparrow R} \psi_{\downarrow R} + \psi_{\uparrow I} \psi_{\downarrow I}\right) \Bigg\} dx.
\end{align}
\end{subequations}%
Although $\mu_\uparrow$ and $\mu_\downarrow$ may initially differ, a true stationary state in the presence of finite Rabi coupling requires convergence to a common value, $\mu_\uparrow = \mu_\downarrow = \mu$.

The corresponding energy functionals are defined as \cite{Fabrelli2021}
\begin{align*}
E_{kin} = & \,\frac{1}{2} \int \left( \left\lvert \frac{\partial \psi_\uparrow}{\partial x} \right\rvert^2 + \left\lvert \frac{\partial \psi_\downarrow}{\partial x} \right\rvert^2 \right) dx,  \\
E_{pot} =  & \int V(x) \left( \lvert\psi_{\uparrow}\rvert^{2} + \lvert\psi_{\downarrow}\rvert^{2} \right) dx, 
\end{align*}
\begin{align*}
E_{int} = & \int \left( \frac{1}{2} g_{\uparrow\uparrow} \lvert\psi_{\uparrow}\rvert^{4} + \frac{1}{2} g_{\downarrow\downarrow} \lvert\psi_{\downarrow}\rvert^{4} + g_{\uparrow\downarrow}\lvert\psi_{\uparrow}\rvert^{2}\lvert\psi_{\downarrow}\rvert^{2} \right) dx,  \\
E_{soc} = & \int k_L \Bigg[ \left(\psi_{\uparrow R} \frac{\partial \psi_{\uparrow I}}{\partial x} - \psi_{\uparrow I} \frac{\partial \psi_{\uparrow R}}{\partial x} \right) \\ & \quad - \left(\psi_{\downarrow R} \frac{\partial \psi_{\downarrow I}}{\partial x} - \psi_{\downarrow I} \frac{\partial \psi_{\downarrow R}}{\partial x} \right) \Bigg] dx, \\
E_{Rab} = & \, 2\Omega \int \left(\psi_{\uparrow R} \psi_{\downarrow R} + \psi_{\uparrow I} \psi_{\downarrow I}\right) dx, 
\end{align*}
where $E_{kin}$, $E_{pot}$, $E_{int}$, $E_{soc}$ and $E_{Rab}$ are the kinetic, potential, interaction, SO coupling, and Rabi coupling energies, respectively. The total energy is given by
\begin{align*}
E = & E_{kin} + E_{pot} + E_{int} + E_{soc} + E_{Rab}.
\end{align*}%
We investigate ground-state excitations in the form of dark--bright solitons in a quasi-one-dimensional harmonic trap, defined by $V(x) = \lambda^2 x^2/2$ with $\lambda = 0.05$. Such a configuration corresponds to the quasi-one-dimensional regime realized under strong transverse confinement, which freezes the radial degrees of freedom while the axial dynamics remain weakly confined~\cite{Pethick2008, Pitaevskii2010, Gerbier2004}. In typical experiments with ultracold alkali gases, including $^{39}$K and $^{87}$Rb condensates, this corresponds to highly anisotropic trapping with $\omega_\perp \gg \omega_x$, where $\omega_\perp$ and $\omega_x$ are the transverse and axial trapping frequencies, respectively. For example, $\omega_\perp \sim 2\pi \times 200$~Hz and $\omega_x \sim 2\pi \times 10$~Hz are consistent with experimentally realiz
ed quasi-one-dimensional geometries~\cite{Goerlitz2001, Moritz2003, DErrico2007, Roati2007}.

For these trapping potential, the natural harmonic oscillator length and time units are $a_{\text{ho}} = \sqrt{\hbar/(m\omega_x)} \approx 3.41\,\mu\text{m}$ and $\omega_x^{-1} \approx 15.9\text{ ms}$ (assuming $m$ for $^{87}\text{Rb}$), respectively. Within this dimensionless framework, a dimensionless Rabi frequency of $\Omega = 1.0$ corresponds to a physical energy scale of $\hbar\Omega = h \times 10\text{ Hz}$. Similarly, a dimensionless spin-orbit coupling strength of $\gamma = 1.0$ maps to a Raman recoil wavevector $k_L = \gamma/a_{\text{ho}} \approx 2.93 \times 10^5\text{ m}^{-1}$, which is readily accessible in laboratory setups by adjusting the intersecting angle of the Raman laser beams~\cite{Lin2011, Sanz2022}.

\section{Reduction of the coupled GP equations to Manakov model}
\label{sec:3}

The Manakov model, introduced by S.~V.~Manakov in 1974~\cite{Manakov1975}, describes the evolution of vector waves in a nonlinear medium where the self-phase-modulation and cross-phase-modulation coefficients are equal, rendering the system completely integrable. The model consists of two coupled one-dimensional nonlinear Schrödinger equations and finds important applications in nonlinear optics (governing polarisation-insensitive soliton transmission in birefringent fibres \cite{Menyuk1988, Evangelides1992}) as well as in the multicomponent BECs with equal intra- and interspecies interactions.

The nature of the supported solitons depends on the sign of the nonlinearity. In the attractive regime (negative nonlinear coefficients), the system supports bright–bright vector solitons that preserve arbitrary polarisation ratios and exhibit rich collision dynamics~\cite{Rajendran2010, Borovkova2011}. In the repulsive regime (positive nonlinear coefficients), dark–dark solitons arise~\cite{Radhakrishnan1995, Radhakrishnan1997, Kevrekidis2016}. Additionally, mixed or multicomponent settings allow for dark–bright solitons, where a localised bright structure is trapped within the effective potential created by a dark soliton background~\cite{Kevrekidis2016}. Such structures also emerge in mixed interaction regimes, where one component is attractive and the other repulsive, leading to co-propagating bound states with a common velocity~\cite{Strecker2002, Ohta2011}.

Owing to its complete integrability, the Manakov system admits a rich hierarchy of exact solutions, including multi-soliton complexes and bound states. These solutions can be systematically constructed using methods such as the inverse scattering transform~\cite{Ablowitz1991}, Hirota’s bilinear formalism~\cite{Hirota2004, Hietarinta1997, Qi2023, Song2020}, and gauge transformation techniques~\cite{Sheppard1997, Rajendran2009, RameshKumar2010}.

In the present context, the coupled Gross--Pitaevskii equations \eqref{eq:gpsoc:1} reduce to the Manakov model in the absence of spin--orbit and Rabi couplings ($k_L \to 0$, $\Omega \to 0$) and external confinement ($V = 0$), provided that all nonlinear interaction coefficients are equal. Under these conditions, the system supports the full family of integrable soliton solutions described above, which serve as fundamental nonlinear excitations revealing key dynamical properties of the model. In what follows, we specifically focus on dark–bright solitons and their stability properties, which will be explored in the subsequent sections.

Beyond this limiting case, our objective is to understand how spin–orbit and Rabi couplings modify the dynamics of such solitonic structures. To this end, we construct a systematic mapping of the full coupled GP system onto the integrable Manakov model in a homogeneous setting ($V=0$). This mapping enables us to employ exact analytical solutions as initial states and subsequently examine how the original coupling terms affect their evolution.

The procedure consists of two consecutive transformations, each designed to remove a specific coupling term. Our starting point is the SO-coupled system of GP equations given in equation \eqref{eq:gpsoc:1}. The first step targets the spin–orbit coupling. To this end, we introduce a Galilean transformation into a co-moving frame \cite{Zhang2012}, defined by
\begin{subequations}
\label{eq:kl}
\begin{align}
\psi_\uparrow = \varphi_{\uparrow} \exp\left[\frac{\mathrm{i}}{2} k_{L}(k_{L}t-2 x)\right], \\
\psi_\downarrow  = \varphi_{\downarrow} \exp\left[\frac{\mathrm{i}}{2} k_{L}(k_{L}t+2 x)\right]. 
\end{align}
\end{subequations}
Physically, this transformation corresponds to imprinting counter-propagating plane-wave phases on the two spinor components. Substituting equation~(\ref{eq:kl}) into the original system, equation~(\ref{eq:gpsoc:1}), eliminates the first-order derivative (SO) terms. The resulting dynamical equations describe a binary condensate with purely Rabi coupling and non-linear interactions:
\begin{subequations}
\label{eq:gpsoc:2}
\begin{align}
\mathrm{i} \partial_{t} \varphi_{\uparrow} = & \left[ -\frac{1}{2}\partial_{x}^2 + g_{\uparrow \uparrow} \lvert \varphi_{\uparrow} \rvert^2 + g_{\uparrow \downarrow} \lvert \varphi_{\downarrow} \rvert^2\right] \varphi_{\uparrow} + \Omega \varphi_{\downarrow}, \label{eq:gpsoc:2a} \\
\mathrm{i} \partial_{t} \varphi_{\downarrow} = & \left[ -\frac{1}{2}\partial_{x}^2 + g_{\downarrow \uparrow} \lvert \varphi_{\uparrow} \rvert^2 + g_{\downarrow \downarrow} \lvert \varphi_{\downarrow} \rvert^2 \right] \varphi_{\downarrow} + \Omega \varphi_{\uparrow}. \label{eq:gpsoc:2b}
\end{align}
\end{subequations}

The second step addresses the remaining Rabi coupling, \( \Omega\), in equations (\ref{eq:gpsoc:2a}) and (\ref{eq:gpsoc:2b}). This term is removed by applying a time-dependent unitary transformation that performs a rotation in the internal spin space:
\begin{align}
\begin{pmatrix}
 \varphi_{\uparrow} \\
 \varphi_{\downarrow}
\end{pmatrix}
 =
\begin{pmatrix}
 \cos\Omega t & -\mathrm{i} \sin \Omega t \\
 -\mathrm{i} \sin \Omega t & \cos\Omega t
\end{pmatrix}
\begin{pmatrix}
 \phi_{\uparrow} \\
 \phi_{\downarrow}
 \end{pmatrix}. \label{eq:rabi}
\end{align}
Implementing this spin rotation diagonalises the linear coupling terms and, under the condition of equal intra- and inter-component interaction strengths, reduces the system to the standard integrable Manakov model \cite{Vinayagam2017, Vinayagam2025, Kartashov2019, Liu2025, Belobo2021}. Having reduced the governing equations to this canonical integrable system, we can directly construct its exact dark-bright soliton solutions. Our analysis employs the dark-bright soliton solution for this system as given in~\cite{Sheppard1997},
\begin{subequations}
\label{eq:db:1}
\begin{align}
\phi_\uparrow = & \tau \left\{ \mathrm{i} \sin \theta + \cos \theta \, \tanh \left[ a \left( x - b t\right) \right] \right\}  \notag \\ & \quad \times \exp\left[ \mathrm{i} c x - \mathrm i \left(\frac{1}{2}c^2 + \tau^2 \right) t \right], \label{eq:db:1a} \\
\phi_\downarrow = & \sqrt{\tau^2 \cos^2\theta - a^2} \, \sech \left[ a \left(x - b t\right) \right]  \notag \\ & \quad \times \exp\left[ \mathrm{i} b x + \mathrm i \left(\frac{1}{2} \left(a^2-b^2 \right) - \tau^2 \right) t \right], \label{eq:db:1b} 
\end{align}
\end{subequations}
where $\theta = -\arctan[(c-b)/a]$. Here, $\tau$ denotes the background amplitude and $c$ determines its phase gradient, while the soliton velocity is governed by the parameter $b$. The soliton width, contrast, and phase jump are controlled through the real parameter $a$, constrained by $a^2 + (c-b)^2 \le \tau^2$ (or $a^2 \le \tau^2 \cos^2 \theta$) \cite{Sheppard1997}.

\subsection{Mass and spin currents of soliton states}
\label{subsec:currents}

To better understand the dynamics and population exchange between the internal states of the condensate, we evaluate the continuity relations for the two-component system. The total mass density $\rho_m = |\psi_\uparrow|^2 + |\psi_\downarrow|^2$ and spin density $\rho_s = |\psi_\uparrow|^2 - |\psi_\downarrow|^2$ satisfy the current conservation equations
\begin{align}
\frac{\partial \rho_{m,s}}{\partial t} + \frac{\partial j_{m,s}}{\partial x} = S_{m,s},
\label{eq:continuity}
\end{align}
where $S_{m,s}$ represent source/sink terms induced by coherent Rabi coupling, and the corresponding mass ($j_m$) and spin ($j_s$) current densities are given by
\begin{align}
j_m(x,t) &= \text{Im}\left[\psi_\uparrow^* \partial_x \psi_\uparrow + \psi_\downarrow^* \partial_x \psi_\downarrow\right] + k_L \left(|\psi_\uparrow|^2 - |\psi_\downarrow|^2\right), \label{eq:mass_current} 
\end{align}
\begin{align}
j_s(x,t) &= \text{Im}\left[\psi_\uparrow^* \partial_x \psi_\uparrow - \psi_\downarrow^* \partial_x \psi_\downarrow\right] + k_L \left(|\psi_\uparrow|^2 + |\psi_\downarrow|^2\right). \label{eq:spin_current}
\end{align}
Since the coherent dynamics conserves the total atom number, the mass density admits no source term, $S_m \equiv 0$, for arbitrary $\Omega$ and $k_L$. Only the spin density has a genuine source, arising solely from Rabi coupling:
\begin{align}
S_s = 4\Omega\,\text{Im}\!\left(\psi_\uparrow^* \psi_\downarrow\right).
\label{eq:spin_source}
\end{align}

\paragraph{Case 1: For $\Omega=0$, $k_L\neq0$.} For this case we find that $S_s = 4\Omega\,\text{Im}(\psi_\uparrow^* \psi_\downarrow) = 0$, which suggest that densities of both the components are locally conserved, i.e, $S_m = S_s = 0$. For the exact stationary dark--bright soliton solution we have
\begin{align}
\psi_{\uparrow,\downarrow} = \phi_{\uparrow,\downarrow}(x)\exp\!\left[\mathrm{i}\left(\frac{k_L^2 t}{2} \mp k_L x\right)\right],
\end{align}
with real envelopes $\phi_{\uparrow,\downarrow}$, one finds $\text{Im}(\psi_\uparrow^* \partial_x \psi_\uparrow) = -k_L \phi_\uparrow^2$ and $\text{Im}(\psi_\downarrow^* \partial_x \psi_\downarrow) = +k_L \phi_\downarrow^2$. Substitution into equations~\eqref{eq:mass_current} and \eqref{eq:spin_current} yields
\begin{subequations}
\begin{align}
j_m &= -k_L \phi_\uparrow^2 + k_L \phi_\downarrow^2 + k_L(\phi_\uparrow^2 - \phi_\downarrow^2) = 0,  \\ 
j_s  & =  -k_L \phi_\uparrow^2 - k_L \phi_\downarrow^2 + k_L(\phi_\uparrow^2 + \phi_\downarrow^2) = 0.
\end{align}
\end{subequations}
Overall we find that both currents vanish identically for this stationary configuration. In particular the phase-gradient contribution exactly cancels the spin--orbit-induced offset. We note that this cancellation is a special property of the stationary ansatz considered here, rather than a generic consequence of setting $\Omega=0$.

\paragraph{Case 2: For $\Omega\neq0$, $k_L=0$.} The spin--orbit contributions to the currents vanish, leaving
\begin{subequations} 
\begin{align}
j_m &= \text{Im}\left[\psi_\uparrow^* \partial_x \psi_\uparrow + \psi_\downarrow^* \partial_x \psi_\downarrow\right], \\ S_m & = 0, \\
j_s &= \text{Im}\left[\psi_\uparrow^* \partial_x \psi_\uparrow - \psi_\downarrow^* \partial_x \psi_\downarrow\right], \\ d S_s & = 4\Omega\,\text{Im}(\psi_\uparrow^* \psi_\downarrow) \neq 0.
\end{align}
\end{subequations}
Here the total mass density is strictly conserved, while the spin density acquires a genuine Rabi-driven source, describing continuous Josephson-like population transfer between the two internal states~\cite{Ravisankar2020_2, Gangwar2022}.

It is worth pointing out that the internal spin dynamics governed by equations~\eqref{eq:continuity}--\eqref{eq:spin_source} can be mapped directly to the Physics of an \emph{internal Josephson junction}~\cite{PhysRevA.59.3890}. Under this scenario, the two hyperfine spin states ($\psi_\uparrow, \psi_\downarrow$) act as two coherent superfluid reservoirs, while the Rabi coupling frequency $\Omega$ plays the role of an effective Josephson tunneling amplitude. By expressing the complex spinor wave functions in amplitude-phase representation, $\psi_{\uparrow,\downarrow}(x,t) = \sqrt{\rho_{\uparrow,\downarrow}(x,t)}\,\mathrm{e}^{\mathrm{i}\phi_{\uparrow,\downarrow}(x,t)}$, the Rabi-induced spin source $S_s$ in equation~\eqref{eq:spin_source} takes the explicit form
\begin{align}
S_s(x,t) = 4\Omega \sqrt{\rho_\uparrow \rho_\downarrow} \sin(\Delta\phi),
\label{eq:josephson_current}
\end{align}
where $\Delta\phi(x,t) = \phi_\downarrow(x,t) - \phi_\uparrow(x,t)$ is the local relative phase between the spin components. Equation~\eqref{eq:josephson_current} directly mirrors the canonical Josephson current-phase relation $I_J = I_c \sin(\Delta\phi)$, with a critical tunneling current density $I_c = 4\Omega \sqrt{\rho_\uparrow \rho_\downarrow}$.

Within this framework, both the dark soliton core and the synthetic gauge field $k_L$ play distinct roles in driving the internal junction dynamics. Across the dark soliton notch, the characteristic phase jump ($\Delta\phi_{\text{core}} \approx \pi$) creates a localized phase gradient. Under non-stationary or perturbed evolution, the time rate of change of the relative phase, $\partial_t(\Delta\phi)$, drives coherent, ac-Josephson-like population exchange between the spin components at frequencies governed by $\Omega$ and the local chemical potential detuning $\Delta\mu = \mu_\uparrow - \mu_\downarrow$. Furthermore, unlike conventional spatial Josephson junctions bounded by physical potential barriers, the Raman-induced spin--orbit coupling $k_L$ imprints a momentum bias that acts as a continuous spatial phase tilt, $\partial_x(\Delta\phi) = -2 k_L$. During dynamical evolution, this gauge-induced shift modulates the local Josephson current-phase relation, giving rise to directionally asymmetric spin-current pulses $j_s(x,t)$ across the soliton structure.

The physical soliton profiles in the original spin–orbit and Rabi-coupled system are obtained by applying the transformations that map the Manakov fields $\phi_j$ to the physical fields $\psi_j$. This yields
\begin{align}\label{eq:sol:SOR}
\psi_{j} = \bigg[\phi_{j} \cos(\Omega t) - \mathrm{i} \phi_{j'} \sin(\Omega t)\bigg]
\exp\left[\mathrm{i}\left(\frac{k^{2}_{L}t}{2}\mp k{_L} x\right)\right],
\end{align}
where $j=\uparrow,\downarrow$ and $j'$ denotes the opposite spin state.

In the limit of vanishing Rabi coupling ($\Omega \to 0$), this reduces to
\begin{align} \label{eq:sol:noR}
\psi_j = \phi_j \exp\left[\mathrm{i}\left(\frac{k_L^2 t}{2} \mp k_L x\right)\right],
\end{align}
whilst in the absence of spin–orbit coupling ($k_L=0$), one obtains
\begin{align}\label{eq:sol:R}
\psi_j = \phi_j \cos(\Omega t) - \mathrm{i}\phi_{j'} \sin(\Omega t).
\end{align}
Whilst the present study focuses on the solution given by equation~\eqref{eq:sol:noR}, the broader objective is to determine the dynamical stability of the full solution described by equation~\eqref{eq:sol:SOR}. The analytical construction provides a convenient starting point for examining how the combined spin–orbit and Rabi couplings influence soliton structure and stability.

Although the sequence of gauge transformations and spin-rotation maps the coupled GP equations \eqref{eq:gpsoc:1} to the integrable Manakov system, this mapping does not preserve the exact solution space when the spin-orbit coupling strength $k_L$ is finite. The origin of this breakdown can be understood from both mathematical and physical perspectives.

First, the gauge transformation used to eliminate the first-order derivative terms introduces spin-dependent phase factors of the form $\exp(\mp i k_L x)$ in the two components. While this transformation successfully removes the explicit derivative coupling at the level of the equations, it simultaneously imparts opposite momenta, $\pm k_L$, to the spin components. As a consequence, the reconstructed wave functions inherently carry a finite relative momentum, $\Delta k = k_\uparrow - k_\downarrow = 2k_L$, which has no counterpart in the standard Manakov model.

The Manakov condensate is characterised by an underlying SU(2) symmetry and supports soliton solutions in which both components share a common velocity and phase gradient. In particular, dark–bright solitons rely on a co-moving structure, where the bright component remains localised within the density dip of the dark component through a precise balance of dispersion and nonlinearity. The presence of a finite relative momentum $\Delta k$ violates this co-moving condition, leading to a continuous dephasing between the components. As a result, the phase coherence required to sustain an exact vector soliton is lost.

\subsection{Impact of Rabi coupling on the dynamics of Manakov soliton}
From a dynamical perspective, the spin–orbit coupling acts as an effective momentum bias or internal current, driving the two components in opposite directions. This induces a relative drift between the dark and bright solitons, which in turn leads to deformation, radiation emission and eventual fragmentation of the structure~\cite{Lin2011, Eiermann2004}. Therefore, even if one considers the Manakov solution as the seed solution, the subsequent time evolution under the original equations deviates from solitonic behaviour when $k_L \neq 0$.

Another way to understand this limitation is through symmetry considerations. While the transformed equations formally resemble the Manakov system, the inverse transformation back to the physical frame reintroduces spin-dependent phase gradients that break the effective Galilean invariance and the degeneracy of momentum states. Consequently, the integrability of the Manakov model does not carry over to the original spin-orbit coupled system, and the exact soliton solutions are no longer supported~\cite{Lin2011, Manakov1975, Sarkar2025}.

The role of Rabi coupling provides additional insight. For finite $\Omega$, the linear coupling between components tends to enforce phase locking, partially compensating for the relative phase mismatch induced by spin–orbit coupling. This can stabilise the soliton structure over intermediate timescales, making the reconstructed solutions appear dynamically robust~\cite{Lin2011, Sarkar2025}. However, in the absence of Rabi coupling ($\Omega = 0$), no such phase synchronisation mechanism exists, and the incompatibility between the imposed momentum difference and the soliton structure manifests itself completely.
\begin{figure}[!htb]
\centering\includegraphics[width=0.99\linewidth]{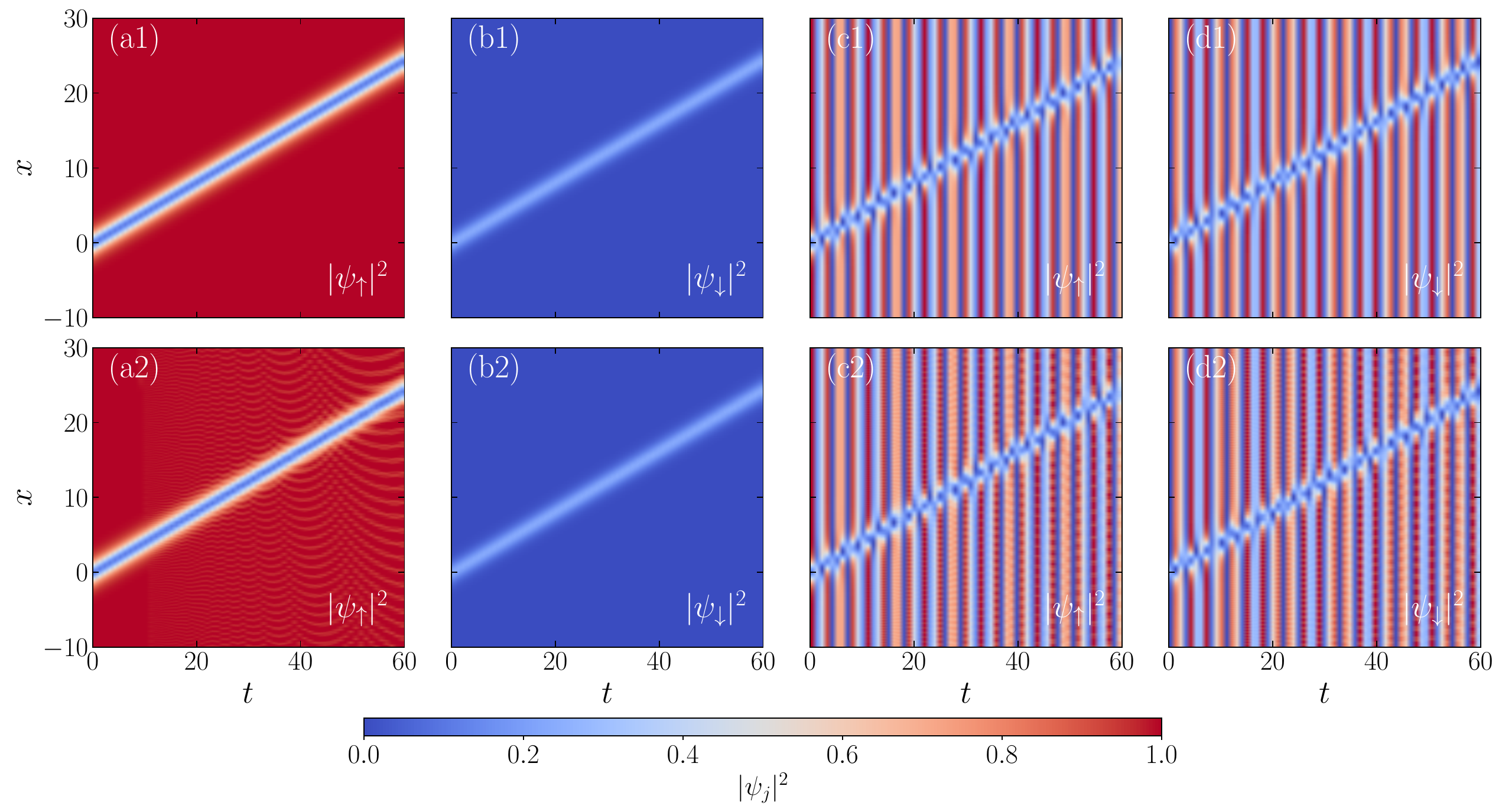}
\caption{Spatiotemporal evolution of component densities $|\psi_{\uparrow}|^2$ and $|\psi_{\downarrow}|^2$ for a dark--bright soliton with $k_L=0$. Top row (a1--d1): exact analytical profiles based on the Manakov solution; bottom row (a2--d2): numerical results via the SSCN method. Left columns (a1,b1 and a2,b2): integrable regime with $\Omega=0$; right columns (c1,d1 and c2,d2): driven regime with $\Omega=2.0$. Soliton parameters in equations \eqref{eq:db:1a} and \eqref{eq:db:1b} are set to $a=0.8$, $b=0.4$, $c=0.1$, and $\tau=1.0$.}
\label{fig:1}
\end{figure}

In contrast, when $k_L = 0$, no spin-dependent phase gradients arise, and an exact mapping to the Manakov model is preserved at both the equation and solution levels. Consequently, the resulting soliton solutions preserve their integrity during time evolution, appearing either as stationary structures for $\Omega = 0$ or as robust, coherently oscillating states for finite $\Omega$. These features are illustrated in figure \ref{fig:1}, which serves as a benchmark for the integrable and near-integrable regimes. The deviations observed for $k_L \neq 0$ in subsequent sections can therefore be directly attributed to the breakdown mechanisms discussed above.

We analyse the dynamics for $k_L = 0$, where the system reduces to a Rabi-coupled binary condensate, to establish a reference for the subsequent study of spin--orbit effects. Figure~\ref{fig:1} compares the spatiotemporal evolution of the component densities obtained from the analytical solutions [Equation \eqref{eq:sol:R}] (top row) with the numerical results produced by our solver (bottom row), both initialised at $t=0$ using the same analytical profiles. In the absence of both spin--orbit and Rabi couplings [$\Omega = 0$ shown in panels (a1)--(b2)], the system reduces to the integrable Manakov model under symmetric interactions. In this limit, the dark--bright soliton propagates without distortion, preserving its shape and velocity. The straight, well-defined trajectories in the $t$--$x$ plane are consistent with the elastic nature of integrable dynamics. The numerical results in panels (a2)--(b2) closely follow the analytical profiles, indicating that the split-Step Crank-Nicolson (SSCN) method \cite{Ravisankar2021} reproduces the expected behaviour.

The introduction of finite Rabi coupling [$\Omega = 2.0$, shown in panels (c1)--(d2)] at $k_L = 0$ breaks integrability and induces coherent inter-component population transfer, visible as temporal modulations in the density evolution~\cite{Lin2011, Sarkar2025, Gangwar2022}. Despite these oscillations, the soliton remains a bound structure, with the dark and bright components staying localised and co-propagating. The numerical results in the second row reproduce these modulations and are consistent with the expected oscillation frequency (of order $\Omega$).

It should be noted that the above discussion holds good for symmetric interactions, which is too restrictive from an experimental point of view, despite preserving the integrability of the system. To delve deep into the dynamics of binary spin-orbit and Rabi coupled condensates in an attempt to bring them closer to experiments, we wish to go beyond symmetric interactions, allowing $k_L$ and $\Omega$ to take finite values. Such an investigation would help us unravel how the localised dark and bright solitons would undergo deformations under the impact of spin-orbit and Rabi coupling. Hence, we now turn to a quantitative characterisation of the system's stability and dynamics. To this end, we formulate the Bogoliubov--de Gennes equations corresponding to the coupled Gross--Pitaevskii model, which provide access to the excitation spectrum and linear stability properties of the stationary states.

In particular, we first obtain stationary solutions via imaginary-time propagation in the absence of Rabi and spin-orbit couplings, both in homogeneous and trapped configurations. The corresponding BdG spectra are then computed to assess their stability and to identify the relevant excitation modes. This spectral analysis serves as a complementary tool to validate the robustness of the numerically obtained states.

Building on this foundation, we proceed to investigate the full nonlinear dynamics by introducing finite $\Omega$ and $k_L$. The subsequent evolution is examined for experimentally relevant regimes with unequal interaction strengths, using initial states constructed from the analytical soliton profiles with additional Gaussian modulation. This facilitates a systematic study of dark-bright soliton dynamics under asymmetric interactions, highlighting how spin-orbit coupling alters their time evolution, besides shedding light on the limiting behaviour in the symmetric interaction regime.

\section{Collective excitation spectrum of dark-bright solitons}
\label{sec:4}

The study of collective excitations is essential for understanding the dynamic properties and stability of solitons in Bose-Einstein condensates. Particularly, dark-bright solitons, which are characterised by a bright component that is embedded within the density dip of a dark soliton background, have garnered significant attention due to their intriguing stability and dynamical behaviors~\cite{Vinayagam2017, Ravisankar2021, Gangwar2023}. The stability of DB solitons in BECs largely depends on the interplay between inter- and intra-species interactions, as well as the external trapping potential. In the repulsive interaction regime, these solitons are generally stable under small perturbations, provided their parameters lie within a permissible range. However, deviations in these parameters, or transitions to different interaction regimes, can induce instabilities or bifurcations in their dynamics, leading to a departure from solitonic behavior~\cite{Katsimiga2018, Xing2025}.

In this section, we investigate the stability characteristics of DB solitons in the presence of spin-orbit and Rabi couplings, as well as under the influence of an external trapping potential. Specifically, we focus on how these couplings affect the collective excitation spectrum of the system, which can provide insight into the stability and dynamics of the solitons.

We begin by perturbing the ground-state wave function and examining the resulting excitation spectrum. This analysis is performed using both analytical and numerical methods, based on the Bogoliubov-de Gennes formalism, which allows us to study the response of the system to small perturbations~\cite{Li2013, Farokh2015, Wang2024a, Xia2023}. The general approach involves introducing a perturbation to the wave function and studying its evolution in time. We then obtain the corresponding eigenfrequencies, which provide information on the collective modes of the system.

The perturbed wave function is written as
\begin{align}
\label{eq:Gro}
\psi_{j}(x, t) = e^{-\mathrm{i}\mu_j t}\left[ \psi_{j}(x,0) + \delta \psi_{j}(x,t) \right],
\end{align}
where $\psi_{j}(x,0)$ denotes the ground-state wave function (at \(t=0\)), with $j \in {\uparrow, \downarrow}$. The perturbation term is taken to be of the form
\begin{align}
\label{eq:Per}
\delta \psi_{j}(x,t) = u_{j}(x)\, \mathrm e^{-\mathrm{i}\nu t} + v^{}_{j}(x)\, \mathrm  e^{\mathrm{i}\nu^{} t}.
\end{align}
where $\nu$ is the frequency of the perturbations, $u_{\uparrow, \downarrow}$ and $v_{\uparrow, \downarrow}$ are the perturbation amplitudes. The Bogoliubov coefficients \( (u_{\uparrow}, u_{\downarrow}, v_{\uparrow}, {v_\downarrow})\) are obtained by substituting (\ref{eq:Gro}) into (\ref{eq:gpsoc:1a}) and (\ref{eq:gpsoc:1b}). The Bogoliubov equation can be written as
\begin{subequations}
\begin{align} 
(\mathcal{M} - \nu \mathrm{I}) (u_{\uparrow}, u_{\downarrow}, v_{\uparrow}, v_{\downarrow})^{T} = 0, \label{eq:bdg}
\end{align}
where
\begin{align}
\mathcal{M} = \begin{pmatrix}
\mathcal{L}_{11}  & \mathcal{L}_{12} & g_{\uparrow\uparrow}\phi_\uparrow^2 & \mathcal{L}_{14} \\
\mathcal{L}_{12}^{*} & \mathcal{L}_{22} & \mathcal{L}_{14} & g_{\downarrow\downarrow}\phi_\downarrow^2 \\
-g_{\uparrow\uparrow}\phi_\uparrow^{*2} & -\mathcal{L}_{14}^{*} & -\mathcal{L}_{11} & -\mathcal{L}_{12}^{*} \\
- \mathcal{L}_{14}^{*} & -g_{\downarrow\downarrow}\phi_\downarrow^{*2} & -\mathcal{L}_{12} & -\mathcal{L}_{22}
\end{pmatrix}
\end{align}
where $\mathrm{I}$ is the $4 \times 4$ unit matrix, and $\mathcal{M}$ is the BdG matrix with
\begin{align}
\mathcal{L}_{11} &= \frac{k^2}{2} - k_L k + V - \mu_1 + 2g_{\uparrow\uparrow}n_\uparrow + g_{\uparrow\downarrow}n_\downarrow, \\
\mathcal{L}_{22} &= \frac{k^2}{2} + k_L k + V - \mu_2 + g_{\downarrow\uparrow}n_\uparrow + 2g_{\downarrow\downarrow}n_\downarrow, \\
\mathcal{L}_{12} &=  \Omega + g_{\uparrow\downarrow}\phi_\uparrow\phi_\downarrow^*, \;\;\;
\text{and}\;\;\; \mathcal{L}_{14} = g_{\uparrow\downarrow} \phi_\uparrow\phi_\downarrow
\end{align}
\end{subequations}
The eigenvalues $\nu$ of this BdG matrix correspond to the frequencies of the collective excitations of the DB solitons. Analysing the excitation spectrum helps us determine the stability of the solitons and the impact of the spin-orbit and Rabi couplings on their dynamics. This formalism also allows us to identify possible instabilities or bifurcations in the soliton solutions, depending on the values of the system parameters.

While the underlying background phase boundaries (such as plane-wave, stripe, and miscible--immiscible transitions) for spin--orbit and Rabi-coupled condensates were mapped in Ref.~\cite{Ravisankar2025}, the BdG spectrum derived here establishes the specific linear stability regimes of the localized dark--bright solitons within these backgrounds.

\section{Quench-induced dynamics of dark–bright solitons}
\label{sec:5}

Building upon the linear stability regimes mapped out via the BdG analysis in section~\ref{sec:4}, we now examine the real-time non-linear dynamics of dark--bright solitons under various physical conditions. As detailed in the following subsections, varying the coupling parameters $(k_L, \Omega)$ and quenching the interaction strengths $g_{ij}$ drives the localized vector solitons from a quiescent bound state into internal breathing modes, and ultimately into modulational-instability-induced multi-soliton fragmentation.

We systematically explore two distinct scenarios: first the Manakov $\mathrm{SU}(2)$-symmetric regime ($g_{\uparrow\uparrow} = g_{\downarrow\downarrow} = g_{\uparrow\downarrow}$) and second the $\mathrm{SU}(2)$-broken asymmetric regime ($g_{\uparrow\uparrow} \neq g_{\downarrow\downarrow} \neq g_{\uparrow\downarrow}$), evaluating both harmonically trapped and homogeneous (untrapped) geometries. In each configuration, stationary dark--bright soliton ground states are first computed via imaginary-time propagation.

Using these prepared states as initial conditions, we perform real-time Gross--Pitaevskii numerical simulations to confirm the dynamical stability of the dark--bright solitons against small perturbations and to track their non-equilibrium evolution under interaction quenches. Throughout this numerical study, the core dark--bright soliton parameters in equations~\eqref{eq:db:1a} and \eqref{eq:db:1b} are held fixed at $a = 1.0$, $b = -0.2$, and $c = 0.1$.

For completeness, we first evaluate the unquenched dynamical evolution of binary BECs hosting Gaussian and dark--bright soliton profiles [equation~\eqref{eq:db:1}]. We begin with the uncoupled scenario ($\Omega = 0$) using equation~\eqref{eq:sol:noR} to isolate synthetic spin--orbit coupling effects, and subsequently incorporate coherent Rabi driving through equation~\eqref{eq:sol:SOR}. This unquenched baseline reveals a rich landscape of coherent phenomena, including stable localized propagation, internal component breathers, interference patterns, second-harmonic generation, and dynamically radiating solitons, thereby setting the stage for the quench-induced dynamics discussed in subsequent subsections.

\subsection{Dynamical Stability of a Binary BEC}

We examine the stability of Gaussian and dark--bright soliton profiles in binary %
\begin{figure*}[!ht]
\includegraphics[width=\linewidth]{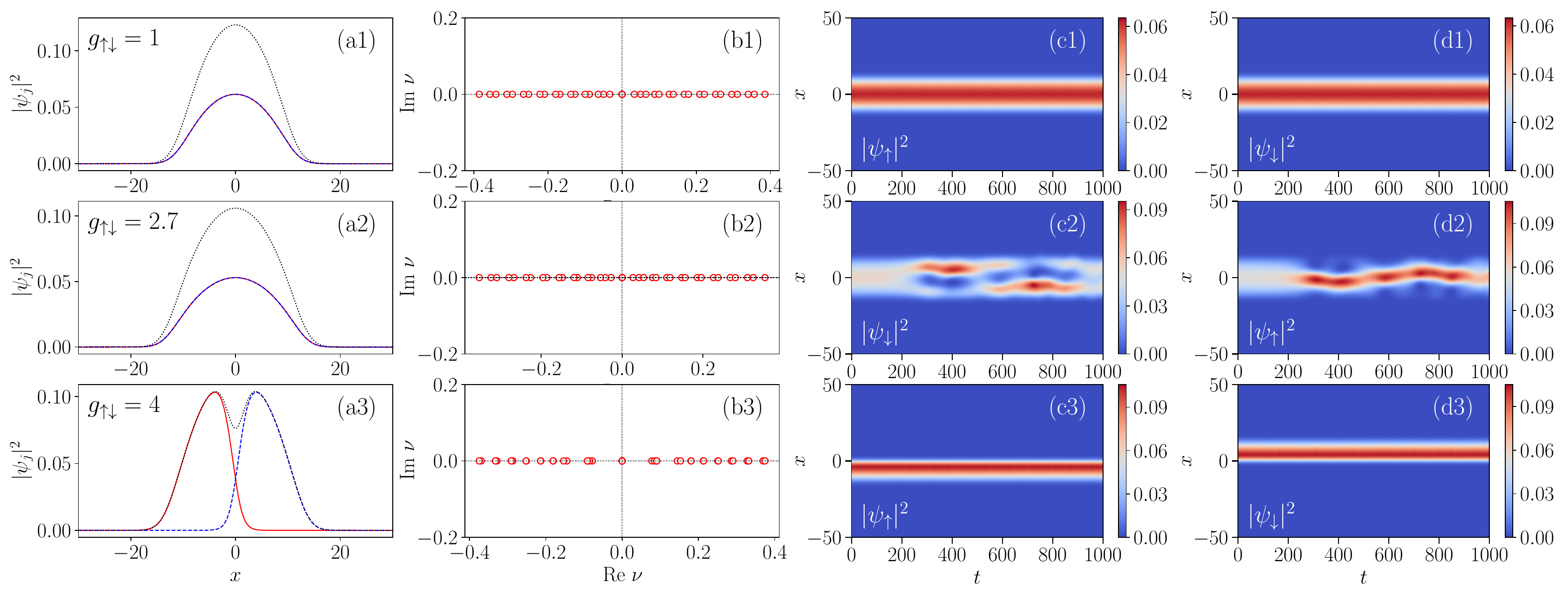}
\caption{Ground-state density profiles $|\psi_\uparrow|^2$ (solid red) and $|\psi_\downarrow|^2$ (dashed blue) with total density (black dotted), prepared via imaginary-time propagation from Gaussian initial states without spin--orbit or Rabi coupling ($k_L = 0$, $\Omega = 0$). Trapped system with $g_{\uparrow\uparrow} = g_{\downarrow\downarrow} = 2$ and $N_\uparrow = N_\downarrow = 1$. Panels (a1)--(a3): ground states for $g_{\uparrow\downarrow} = g_{\downarrow\uparrow} = 1$, $2.7$, and $4$, with chemical potentials $\mu_\uparrow = \mu_\downarrow = 0.1878$, $0.2517$, and $0.2414$, respectively. Panels (b1)--(b3): corresponding BdG excitation spectra. Panels (c1)--(d3): real-time evolution of $|\psi_\uparrow|^2$ and $|\psi_\downarrow|^2$ under small perturbations.}
\label{Asymmetric-main/Ground-WT-WOT-PW}
\end{figure*}
\begin{figure*}[!ht]
\includegraphics[width=\linewidth]{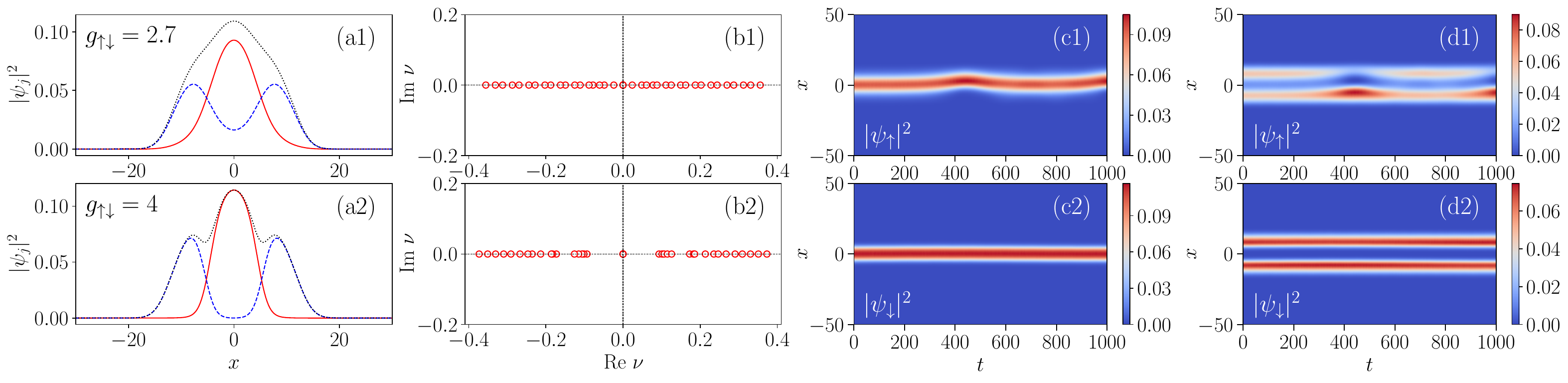}
\caption{ Ground-state density profiles $|\psi_\uparrow|^2$ and $|\psi_\downarrow|^2$ prepared via imaginary-time propagation using Gaussian states superimposed with dark--bright soliton profiles from equation~\eqref{eq:db:1} ($k_L = 0$, $\Omega = 0$). Trapped system with $g_{\uparrow\uparrow} = g_{\downarrow\downarrow} = 2$ and $N_\uparrow = N_\downarrow = 1$. Panels (a1),(a2): ground states for $g_{\uparrow\downarrow} = g_{\downarrow\uparrow} = 2.7$ and $4.0$, with chemical potentials $(\mu_\uparrow, \mu_\downarrow) = (0.2418, 0.2507)$ and $(0.2385, 0.2642)$, respectively. Panels (b1),(b2): corresponding BdG excitation spectra. Panels (c1)--(d2): real-time evolution of $|\psi_\uparrow|^2$ and $|\psi_\downarrow|^2$ under small perturbations.}
\label{Asymmetric-main/Ground-WT-WOT-DB}
\end{figure*}
Bose--Einstein condensates, considering both harmonically trapped and untrapped configurations.  We initially set $k_L = 0$ and $\Omega = 0$ during both imaginary- and real-time propagation to establish a baseline for interaction effects without synthetic spin--orbit or Rabi couplings. We consider a partially symmetric interaction scenario where intraspecies coupling strengths are equal ($g_{\uparrow\uparrow} = g_{\downarrow\downarrow} = 2.0$), while interspecies interaction $g_{\uparrow\downarrow} = g_{\downarrow\uparrow}$ is varied. The normalisation is fixed at $N_\uparrow = N_\downarrow = 1$.

In figure~\ref{Asymmetric-main/Ground-WT-WOT-PW}, we present the ground-state density profiles and subsequent dynamical evolution for interspecies interaction strengths $g_{\uparrow\downarrow} = 1$, $2.7$, and $4.0$ under harmonic confinement ($\lambda = 0.05$). Ground states are obtained via imaginary-time propagation starting from Gaussian initial wave functions. As $g_{\uparrow\downarrow}$ increases, the system transitions from miscible to immiscible configurations, governed by the miscibility parameter
\begin{align}
\label{eq:Phase_transition}
\Delta = \frac{g_{\uparrow\downarrow}^2}{g_{\uparrow\uparrow} g_{\downarrow\downarrow}}.
\end{align}
For $g_{\uparrow\uparrow} = g_{\downarrow\downarrow} = 2.0$, the miscibility threshold is $g_{\uparrow\downarrow}^c = 2.0$. For $g_{\uparrow\downarrow} = 1.0$ ($\Delta = 0.25$), the component densities overlap significantly, forming a miscible state [figure~\ref{Asymmetric-main/Ground-WT-WOT-PW}(a1)]. For $g_{\uparrow\downarrow} = 2.7$ and $4.0$ ($\Delta > 1$), the components undergo spatial separation into immiscible states [figures~\ref{Asymmetric-main/Ground-WT-WOT-PW}(a2) and \ref{Asymmetric-main/Ground-WT-WOT-PW}(a3)]. The corresponding chemical potentials are $\mu_\uparrow = \mu_\downarrow = 0.1878$, $0.2517$, and $0.2414$ for panels (a1), (a2), and (a3), respectively.

The linear stability of these stationary states is confirmed by the Bogoliubov--de Gennes (BdG) excitation spectra in panels (b1)--(b3) of figure~\ref{Asymmetric-main/Ground-WT-WOT-PW}, which display purely real eigenfrequencies. The corresponding real-time evolution following small perturbations is shown in panels (c1)--(d3). While the miscible ($g_{\uparrow\downarrow} = 1$) and strongly immiscible ($g_{\uparrow\downarrow} = 4$) states exhibit robust, bounded oscillations, the state at $g_{\uparrow\downarrow} = 2.7$ [figure~\ref{Asymmetric-main/Ground-WT-WOT-PW}(a2)] represents a metastable configuration with symmetric chemical potentials ($\mu_\uparrow = \mu_\downarrow = 0.2517$). Under real-time propagation, small perturbations break this spatial symmetry, causing the components to phase-separate dynamically as seen in figures~\ref{Asymmetric-main/Ground-WT-WOT-PW}(c2) and \ref{Asymmetric-main/Ground-WT-WOT-PW}(d2).

In the absence of external confinement ($\lambda = 0$), imaginary-time evolution from Gaussian states converges to the trivial zero-density solution. To obtain non-trivial homogeneous states, localized profiles are evolved without the trap. In this untrapped setting, miscible configurations spread gradually, whereas immiscible states form domain-wall-like structures that maintain localized interfaces during real-time propagation.

We next analyze the dark--bright (DB) soliton solution [equation~\eqref{eq:db:1}] under harmonic confinement, as depicted in figure~\ref{Asymmetric-main/Ground-WT-WOT-DB}. The ground-state density profiles obtained via imaginary-time propagation superimposed with dark--bright soliton initial guesses are shown in panels (a1) for $g_{\uparrow\downarrow} = 2.7$ and (a2) for $g_{\uparrow\downarrow} = 4.0$, with corresponding chemical potentials $(\mu_\uparrow, \mu_\downarrow) = (0.2418, 0.2507)$ and $(0.2385, 0.2642)$, respectively. In both cases, the dark--bright soliton forms a stable, localized composite structure where the dark component occupies the center and the bright component is localized within the notch.

The linear stability of these dark--bright soliton states is established by the BdG spectra in figures~\ref{Asymmetric-main/Ground-WT-WOT-DB}(b1) and \ref{Asymmetric-main/Ground-WT-WOT-DB}(b2), which consists entirely of real eigenfrequencies. Long-term real-time dynamics, shown in figures~\ref{Asymmetric-main/Ground-WT-WOT-DB}(c1)--(d2), confirm that the dark--bright solitons undergo stable, periodic breathing oscillations without structural decay, demonstrating their robustness under harmonic confinement prior to the introduction of spin--orbit and Rabi couplings.

\subsection{Effect of SO and Rabi coupling on the dynamics of Dark-Bright solitons.} 
\label{sec:5-A(2)}

To further investigate the role of spin–orbit (SO) coupling, we consider dark–bright soliton solutions in the presence of the $k_{L}$ term, as introduced in equation~(\ref{eq:kl}). These solutions satisfy the coupled Gross–Pitaevskii equations and allow us to systematically analyse the combined influence of SO and Rabi coupling on the soliton dynamics.
\begin{figure}[!htb]
\centering\includegraphics[width=0.99\linewidth]{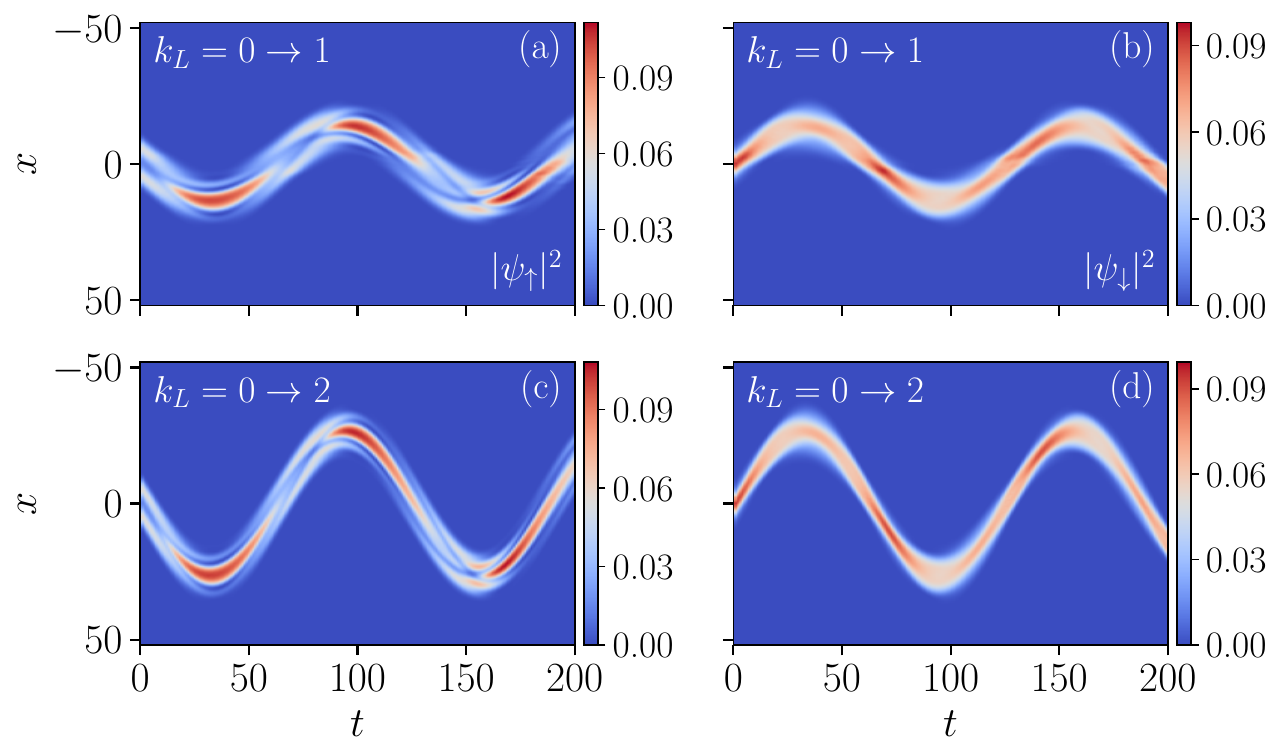}
\caption{Time evolution of dark-bright soliton densities inside a harmonic trap under varying spin--orbit coupling strength. Panels (a) and (b) show $|\psi_{\uparrow}|^2$ and $|\psi_{\downarrow}|^2$, respectively, for $k_L$ increasing from $0$ to $1$ with $\Omega = 0$. Panels (c) and (d) show $|\psi_{\uparrow}|^2$ and $|\psi_{\downarrow}|^2$, respectively, for $k_L$ increasing from $0$ to $2$ with $\Omega = 0$. The nonlinearity parameters are fixed at $g_{\uparrow\uparrow} = g_{\downarrow\downarrow} = 2.0$ and $g_{\uparrow\downarrow} = g_{\downarrow\uparrow} = 2.7$.}
\label{Asymmetric-main/New_Coupling_2_2_2p7_Kl1_Kl2}
\end{figure}
The dark–bright soliton is employed as the initial condition to obtain the ground state via imaginary-time propagation, in the absence of the SO coupling term ($k_{L}=0$). Subsequently, the SO coupling is introduced, while keeping the Rabi coupling switched off, to examine the real-time evolution of the solitons. Thus, the SO coupling is incorporated only during real-time propagation to analyse its effect on the soliton dynamics. We consider two scenarios: one in the presence of an external trapping potential and the other without a trap. The nonlinear interaction strengths are chosen as $g_{\uparrow \uparrow} = g_{\downarrow \downarrow} = 2.0$ and $g_{\uparrow \downarrow} = g_{\downarrow \uparrow} = 2.7$.

In figure~\ref{Asymmetric-main/New_Coupling_2_2_2p7_Kl1_Kl2} we show the time evolution of dark–bright soliton densities confined in a harmonic trap for different strengths of spin–orbit coupling. The time-evolution density profiles displayed in figure~\ref{Asymmetric-main/New_Coupling_2_2_2p7_Kl1_Kl2}(a) for $|\psi_{\uparrow}|^2$ and (b) for $|\psi_{\downarrow}|^2$ correspond to the case when the SO coupling is quenched from $k_L=0$ to $k_{L}=1$ for $\Omega=0$. In this regime, the densities exhibit relatively narrow and well-defined oscillations, indicating localised but dynamically active structures. 
As the spin–orbit (SO) coupling strength is increased from $k_{L}=1$ to $k_{L}=2$, a clear broadening of these oscillations is observed, as shown in panels (c) for $|\psi_{\uparrow}|^2$ and (d) for $|\psi_{\downarrow}|^2$. This broadening suggests an enhancement in the spatial extent of the oscillatory motion. These observations demonstrate that SO coupling alone is sufficient to generate oscillatory dynamics in both spin components, while the underlying soliton profiles remain robust and dynamically stable throughout the evolution.

\begin{figure}[!htb]
\centering\includegraphics[width=0.99\linewidth]{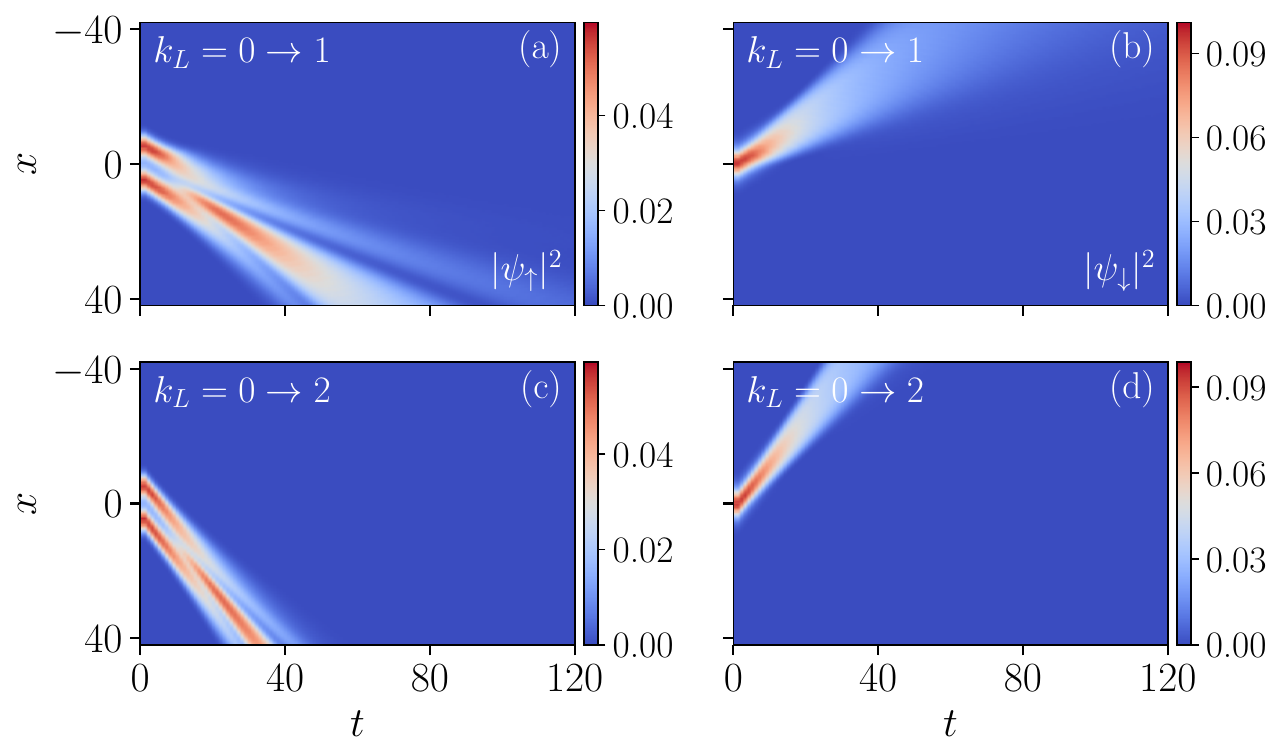}
\caption{Time evolution of dark-bright soliton densities in the absence of a harmonic trap. Panels (a) and (b) show $|\psi_{\uparrow}|^2$ and $|\psi_{\downarrow}|^2$, respectively, for $k_L$ increasing from $0$ to $1$ with $\Omega = 0$. Panels (c) and (d) show $|\psi_{\uparrow}|^2$ and $|\psi_{\downarrow}|^2$, respectively, for $k_L$ increasing from $0$ to $2$ with $\Omega = 0$. The nonlinearity parameters are fixed at $g_{\uparrow\uparrow} = g_{\downarrow\downarrow} = 2.0$ and $g_{\uparrow\downarrow} = g_{\downarrow\uparrow} = 2.7$.}
\label{Asymmetric-main/New_Coupling_WOT_2_2_2p7_Kl1_Kl2}
\end{figure}

In figure~\ref{Asymmetric-main/New_Coupling_WOT_2_2_2p7_Kl1_Kl2} we show the time-evolution density profiles for dark-bright soliton once the SO coupling is quenched from $k_L=0\to1$ (a, b) and $k_L=0\to2$ (c,d), for $|\psi_{\uparrow}|^2$ and $|\psi_{\downarrow}|^2$ respectively at $\Omega=0$, which reveal a qualitatively different behavior that those without the trapping potential. In this case, the two spin components propagate in opposite directions: the $|\psi_{\uparrow}|^{2}$ density shifts toward the positive direction, whereas the $|\psi_{\downarrow}|^{2}$ density moves toward the negative direction. Upon increasing the SO coupling strength to $k_{L}=2$, this counter-propagating behaviour persists, as illustrated in panels (c) and (d), although noticeable modifications in the amplitudes of the solitons become apparent. This indicates that while the directionality is preserved, the strength of the SO coupling influences the amplitude and possibly the dispersion characteristics of the wave packets.

\begin{figure}[!htb]
\includegraphics[width=0.99\linewidth]{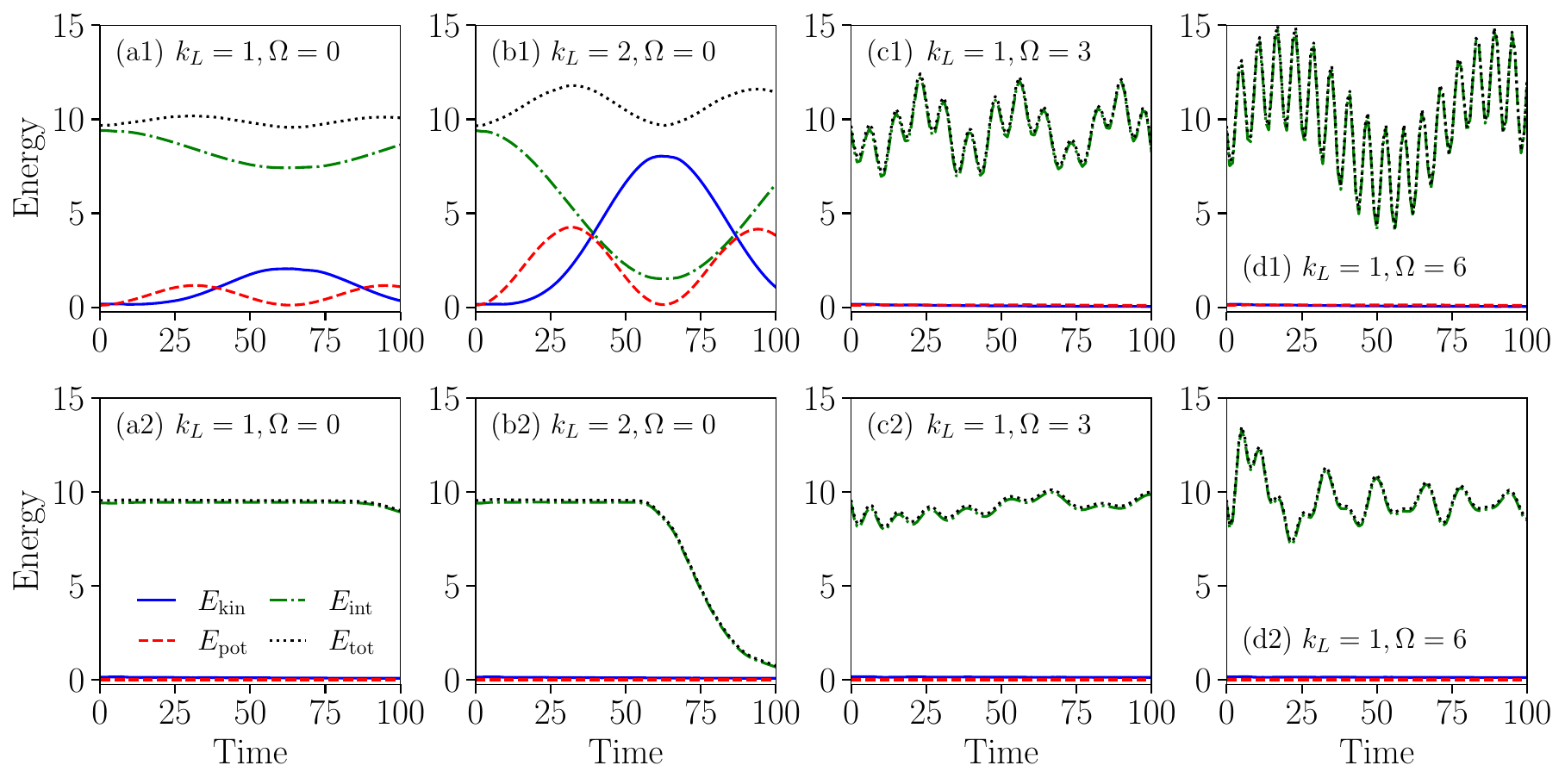}
\caption{Time evolution of various energy components---kinetic ($E_{\mathrm{kin}}$), potential ($E_{\mathrm{pot}}$), interaction ($E_{\mathrm{int}}$), and total ($E_{\mathrm{tot}}$)---during imaginary- and real-time propagation, both in the presence (a1)-(d1) and absence (a2)-(d2) of a harmonic trap. The non-linearity parameters are fixed at $g_{\uparrow\uparrow} = g_{\downarrow\downarrow} = 2.0$ and $g_{\uparrow\downarrow} = g_{\downarrow\uparrow} = 2.7$. The coupling parameters for each panel are: (a1) and (a2): $k_L = 1$, $\Omega = 0$; (b1) and (b2): $k_L = 2$, $\Omega = 0$; (c1) and (c2): $k_L = 1$, $\Omega = 3$; (d1) and (d2): $k_L = 1$, $\Omega = 6$.}
\label{Asymmetric-main/Energy_Asymmetric-WOC-IM}
\end{figure}

To gain deeper insight into the role of different energy contributions in governing the overall dynamics of the condensate, we analyze the temporal evolution of the kinetic ($E_{\mathrm{kin}}$), potential ($E_{\mathrm{pot}}$), interaction ($E_{\mathrm{int}}$), and total energy ($E_{\mathrm{tot}}$), as presented in figure~\ref{Asymmetric-main/Energy_Asymmetric-WOC-IM}. In particular, the upper panels [(a1)–(d1)] correspond to the system in the presence of a harmonic trap, while the lower panels [(a2)–(d2)] represent the trapless case, allowing for a clear comparison of confinement effects. We first consider the dynamics in the absence of the trap. For $k_{L}=1$ and $\Omega=0$ [figure~\ref{Asymmetric-main/Energy_Asymmetric-WOC-IM}(a2)], all three energy components remain essentially non-oscillatory over time, indicating a dynamically stable regime with negligible excitation. The total energy is nearly conserved, exhibiting only a slight monotonic decrease, which may arise from weak numerical dissipation. A qualitatively similar behaviour is observed for $k_{L}=2$ and $\Omega=0$ [figure~\ref{Asymmetric-main/Energy_Asymmetric-WOC-IM}(b2)], although in this case the total energy decays somewhat faster, reflecting the enhanced influence of stronger spin–orbit coupling on the system’s energetics. Upon introducing finite Rabi coupling, the dynamics change significantly. For $k_{L}=1$ and $\Omega=3$ in the absence of the trap [figure~\ref{Asymmetric-main/Energy_Asymmetric-WOC-IM}(c2)], both $E_{\mathrm{kin}}$ and $E_{\mathrm{pot}}$ remain negligible, while the interaction energy exhibits sustained, non-decaying oscillations driven by the interplay between spin–orbit and Rabi couplings. Despite these oscillations, the total energy remains conserved up to small bounded fluctuations, confirming the stability of the dynamics. In the presence of the harmonic trap [figure~\ref{Asymmetric-main/Energy_Asymmetric-WOC-IM}(c1)], a similar trend persists, with $E_{\mathrm{kin}}$ and $E_{\mathrm{pot}}$ remaining small and $E_{\mathrm{int}}$ showing regular oscillations, likely arising from coherent population transfer between spin components induced by the Rabi coupling, while $E_{\mathrm{tot}}$ remains well conserved. Increasing the Rabi coupling further to $\Omega=6$ while keeping $k_{L}=1$ leads to a noticeable increase in the oscillation frequency of the interaction energy, as seen in both the trapped [figure~\ref{Asymmetric-main/Energy_Asymmetric-WOC-IM}(d1)] and trapless [figure~\ref{Asymmetric-main/Energy_Asymmetric-WOC-IM}(d2)] cases. This reflects the stronger coupling-induced quench in the system. Although the oscillations become more rapid and pronounced, the total energy remains conserved in both scenarios, demonstrating the robustness and dynamical stability of the condensate even under strong coupling conditions.

Next, we study the dark–bright soliton dynamics under the combined effects of SO and Rabi coupling as given in (\ref{eq:kl}) and (\ref{eq:rabi}), both in the presence and absence of a trap. As in the previous section, the initial ground state is prepared without SO or Rabi coupling using imaginary-time propagation. Subsequently, the SO and Rabi coupling terms are introduced during real-time evolution to investigate their influence on the soliton dynamics. The non-linear interaction strengths are chosen $g_{\uparrow \uparrow} = g_{\downarrow \downarrow} = 2.0$, $g_{\uparrow \downarrow} = g_{\downarrow \uparrow} = 2.7$.

\begin{figure}[!htb]
\centering\includegraphics[width=0.99\linewidth]{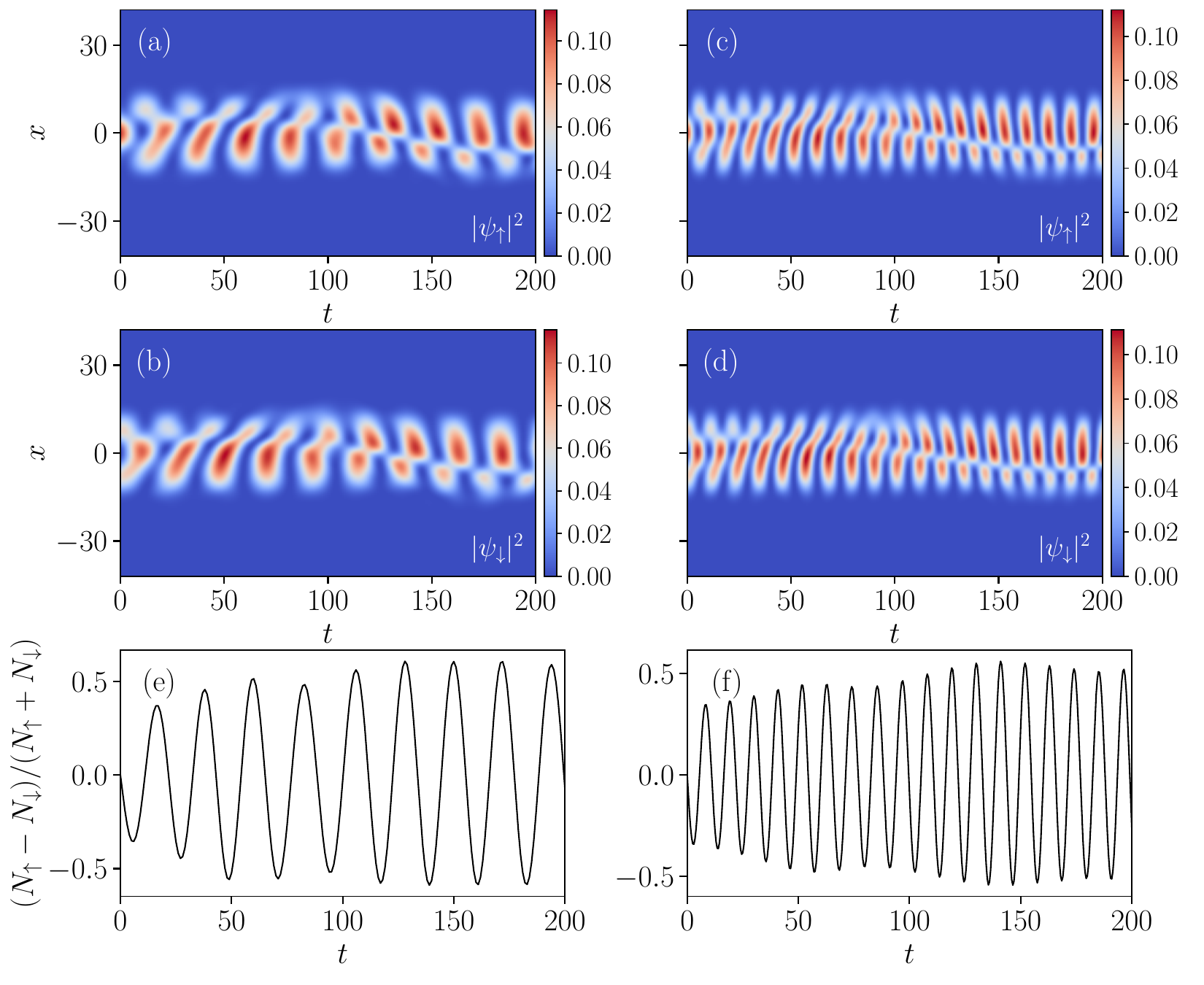}
\caption{Time evolution of dark–bright soliton densities in a harmonically trapped spin–orbit-coupled Bose–Einstein condensate.  Panels (a) and (b) show the component densities $|\psi_{\uparrow}|^2$ and $|\psi_{\downarrow}|^2$, respectively, following a quench of the Rabi frequency from $\Omega = 0$ to $\Omega = 3$, with spin–orbit coupling strength $k_L = 1$.  Panels (c) and (d) present the corresponding dynamics for a stronger quench, $\Omega = 0 \to 6$, under identical conditions. The interaction parameters are fixed at $g_{\uparrow\uparrow} = g_{\downarrow\downarrow} = 2.0$ and $g_{\uparrow\downarrow} = g_{\downarrow\uparrow} = 2.7$.  Panels (e) and (f) display the associated population imbalance, $(N_\uparrow - N_\downarrow)/(N_\uparrow + N_\downarrow)$, as a function of time for the quenches $\Omega: 0 \to 3$ and $\Omega: 0 \to 6$, respectively.}
\label{Asymmetric-main/New_C1_TE_Om3_Om6_Kl1}
\end{figure}

Figure~\ref{Asymmetric-main/New_C1_TE_Om3_Om6_Kl1} illustrates the real-time evolution of dark–bright soliton structures in a harmonically trapped spin–orbit-coupled Bose–Einstein condensate following a sudden quench of the Rabi coupling. Panels (a) and (b) depict the spatiotemporal dynamics of the component densities $|\psi_{\uparrow}|^{2}$ and $|\psi_{\downarrow}|^{2}$, respectively, for a moderate quench from $\Omega=0$ to $\Omega=3$ at fixed spin–orbit coupling strength $k_{L}=1$. The initial stationary configuration evolves into a dynamically robust dark–bright soliton, where a density dip (dark soliton) in one component is accompanied by a localised density peak (bright soliton) in the other. The soliton exhibits coherent oscillatory motion within the harmonic trap, maintaining its integrity over long times, which indicates dynamical stability under moderate coupling. Panels (c) and (d) show the corresponding evolution for a stronger quench from $\Omega=0$ to $\Omega=6$. In this case, the soliton dynamics become more pronounced, with faster oscillations and enhanced density modulations, reflecting the stronger intercomponent coupling induced by the larger Rabi frequency. Despite the increased dynamical activity, the dark–bright soliton structure remains intact, demonstrating robustness against stronger quenches. Panels (e) and (f) present the time evolution of the population imbalance, $(N_{\uparrow}-N_{\downarrow})/(N_{\uparrow}+N_{\downarrow})$, corresponding to the two quench protocols. For $\Omega:0\rightarrow3$, the imbalance exhibits regular, nearly sinusoidal oscillations, indicating coherent population exchange between the spin components. For the stronger quench $\Omega:0\rightarrow6$, the oscillations become more rapid, consistent with the increased Rabi coupling strength, while remaining well-defined, further confirming the coherent and stable nature of the dynamics. The periodic nature of these oscillations suggests a coherent exchange of population between the two spin components, with the Rabi coupling acting as the driving force for the transfer. As the Rabi coupling strength increases, the frequency of the oscillations also increases, reflecting the stronger coupling and faster population dynamics between the spin-up and spin-down components~\cite{Ravisankar2020_2, Gangwar2022, Gangwar2023}.

\begin{figure}[!htb] 
\centering\includegraphics[width=1.0\linewidth]{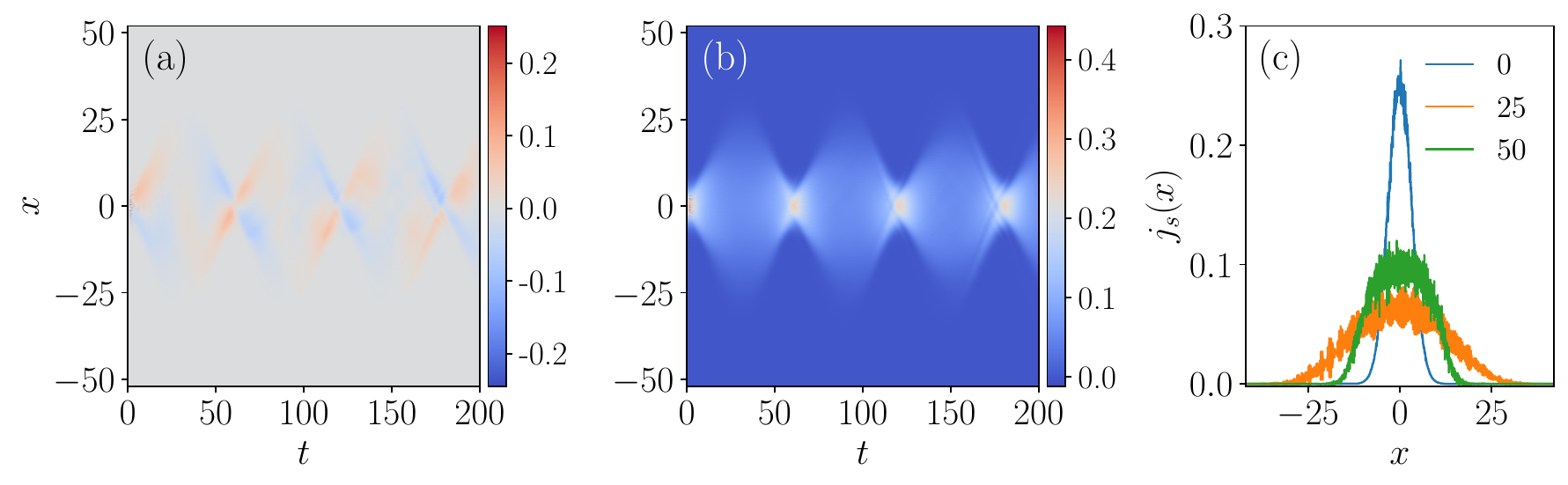} 
\caption{ Spatiotemporal evolution of (a) mass current density $j_m(x,t)$ and (b) spin current density $j_s(x,t)$ for a dark--bright soliton in a harmonic trap with $k_L = 1.0$ and $\Omega = 3.0$. Panel (c) shows spatial profiles of $j_s(x)$ at representative time snapshots ($t = 0, 25, 50$). The interaction parameters are $g_{\uparrow\uparrow} = g_{\downarrow\downarrow} = 2.0$ and $g_{\uparrow\downarrow} = 2.7$.}
\label{fig:currents} 
\end{figure}

To quantitatively track internal transport and gauge-induced flow, we evaluate the real-time evolution of the mass current density $j_m(x,t)$ equation~(\ref{eq:mass_current}) and spin current density $j_s(x,t)$ equation~(\ref{eq:spin_current}). As shown in figures~\ref{fig:currents}(a) and \ref{fig:currents}(b) for $k_L = 1.0$ and $\Omega = 3.0$, both current distributions exhibit a repeating diagonal, X-shaped pattern arising from the periodic oscillation of the soliton within the harmonic trap. Maxima in current magnitude occur as the soliton traverses the trap center, while minima coincide with its turning points.

The spin current in figure~\ref{fig:currents}(b) exhibits a significantly larger overall magnitude than the mass current. This is due to the static spin-orbit background contribution, $-k_L\rho_m = -k_L \left( |\psi_\uparrow|^2+| \psi_\downarrow|^2 \right)$, superposed on the oscillatory phase-gradient term. Superimposed on this trap-induced trajectory is a faster modulation dictated by the Rabi frequency $\Omega$, corresponding to coherent, Josephson-like population exchange between the two spin components. Figure~\ref{fig:currents}(c) displays the spatial profiles of $j_s(x)$ at representative times ($t = 0, 25, 50$): the sharp, narrow peak at $t=0$ marks the soliton's passage through the trap center, whereas the broader, modulated profiles at $t=25$ and $t=50$ capture the interplay between the slower kinematic motion of the soliton and the faster internal spin dynamics driven by $\Omega$.

Thus, while the mass current $j_m(x,t)$ displays bounded, localized oscillations tracking the center-of-mass motion of the vector soliton, the spin current density $j_s(x,t)$ simultaneously encodes two distinct mechanisms: the trap-driven trajectory and the Rabi-induced Josephson population transfer across the soliton notch, superimposed on the persistent gauge offset $-k_L\rho_m$.

\begin{figure}[!htb]
\centering\includegraphics[width=0.99\linewidth]{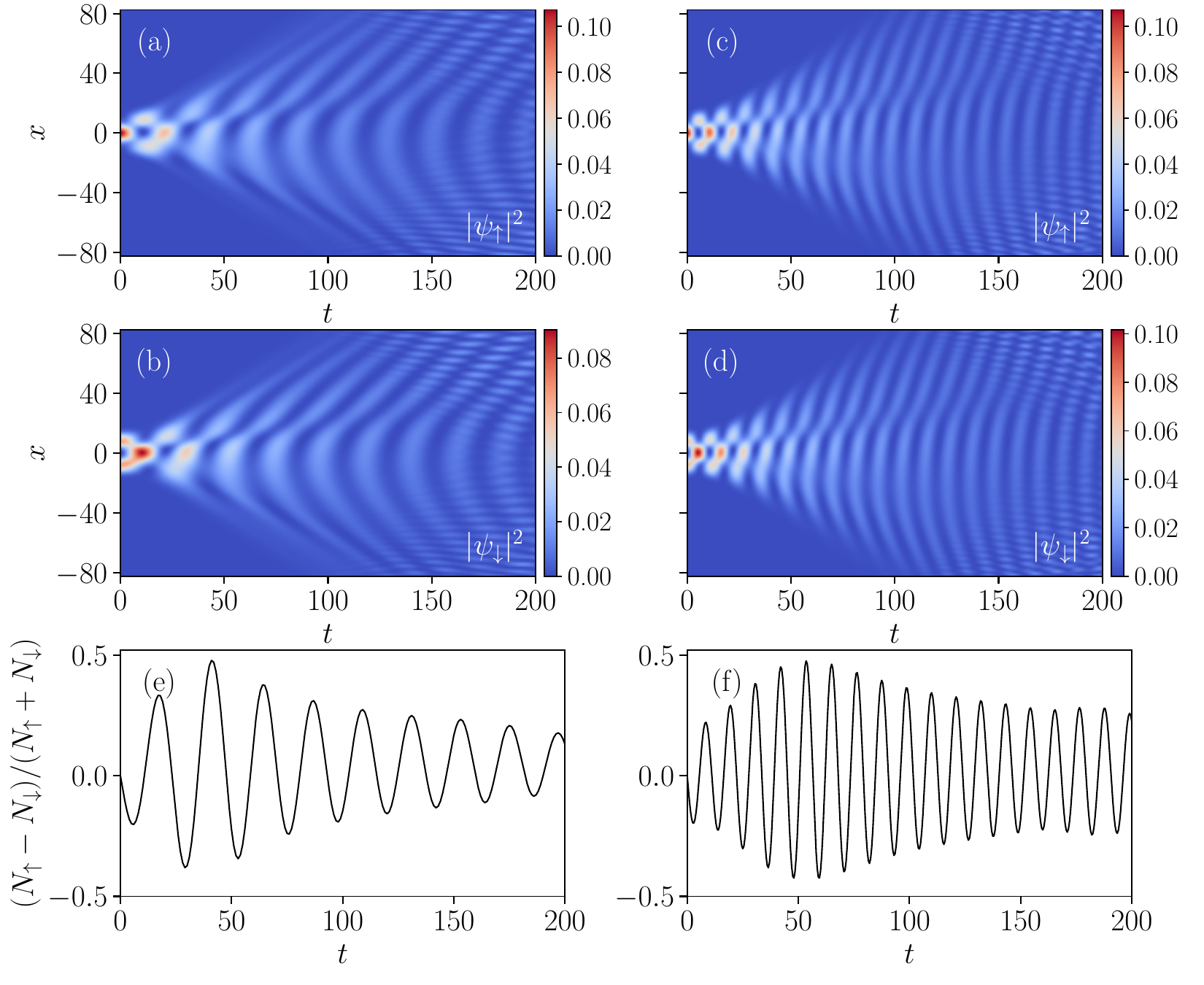}
\caption{
Time evolution of dark--bright soliton densities in a spin--orbit-coupled Bose--Einstein condensate in the absence of an external harmonic trap.  Panels (a) and (b) show the component densities $|\psi_{\uparrow}|^2$ and $|\psi_{\downarrow}|^2$, respectively, following a quench of the Rabi frequency from $\Omega = 0$ to $\Omega = 3$, with the spin--orbit coupling strength varied from $k_L = 0$ to $k_L = 1$.  Panels (c) and (d) present the corresponding dynamics for a stronger quench, $\Omega = 0 \to 6$, at fixed $k_L = 1$.  The interaction parameters are fixed at $g_{\uparrow\uparrow} = g_{\downarrow\downarrow} = 2.0$ and $g_{\uparrow\downarrow} = g_{\downarrow\uparrow} = 2.7$.  Panels (e) and (f) display the associated population imbalance, $(N_\uparrow - N_\downarrow)/(N_\uparrow + N_\downarrow)$, as a function of time for the quenches $\Omega: 0 \to 3$ and $\Omega: 0 \to 6$, respectively.
}
\label{Asymmetric-main/WOC-WOT-Kl1Om3-Kl1Om6}
\end{figure}%
\begin{figure*}[!htb]
\includegraphics[width=0.99\linewidth]{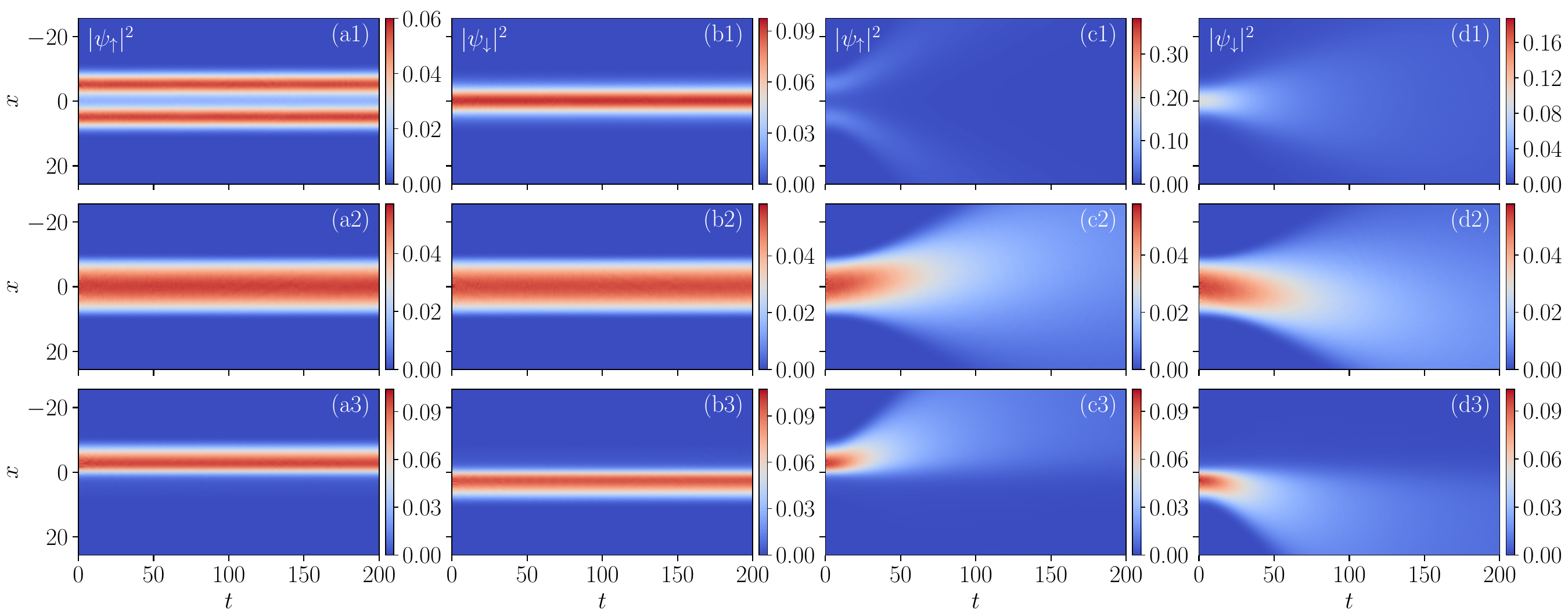}
\caption{ Spatiotemporal evolution of component densities $|\psi_{\uparrow}|^2$ and $|\psi_{\downarrow}|^2$ for dark--bright solitons under varying coupling parameters $(k_L,\Omega)$. First and second columns: trapped system ($\lambda = 0.05$); third and fourth columns: untrapped system ($\lambda = 0$). Rows correspond to coupling parameter sets: (a1)--(d1) $(1,0)$, (a2)--(d2) $(2,5)$, and (a3)--(d3) $(4,5)$. For each configuration, upper and lower panels show spin-up and spin-down components, respectively.}
\label{new:fig:9}
\end{figure*}

Motivated by the above results in the presence of harmonic confinement, it is instructive to examine how the absence of an external trapping potential modifies the soliton dynamics and coherence properties of the system. In particular, removing the trap allows us to isolate the intrinsic role of spin–orbit and Rabi couplings in governing the evolution of dark–bright soliton structures. Figure~\ref{Asymmetric-main/WOC-WOT-Kl1Om3-Kl1Om6} presents the corresponding real-time dynamics in the trapless case. Panels (a) and (b) show the evolution of the component densities $|\psi_{\uparrow}|^{2}$ and $|\psi_{\downarrow}|^{2}$ following a quench of the Rabi frequency from $\Omega=0$ to $\Omega=3$, where the spin–orbit coupling strength is varied from $k_{L}=0$ to $k_{L}=1$. In contrast to the trapped case, the absence of confinement leads to a more pronounced spreading of the condensate background, although the dark–bright soliton structure remains clearly identifiable. The inclusion of finite spin–orbit coupling modifies the propagation characteristics of the soliton, leading to asymmetric density profiles and altered velocity, highlighting the crucial role of $k_{L}$ in shaping the dynamics.
For a stronger quench, $\Omega:0\rightarrow6$, as shown in panels (c) and (d), the system exhibits significantly enhanced dynamical activity. The soliton undergoes faster evolution with more prominent density modulations and partial dispersive spreading, reflecting the stronger intercomponent coupling. Nevertheless, the core dark–bright soliton structure persists over the timescale considered, indicating robustness even in the absence of external confinement. Panels (e) and (f) display the corresponding population imbalance dynamics for the two quench protocols. For $\Omega:0\rightarrow3$, the imbalance shows coherent oscillations, albeit with slight amplitude modulation due to the absence of trapping. In the case of the stronger quench $\Omega:0\rightarrow6$, the oscillations become more rapid and exhibit increased complexity, consistent with the enhanced coupling strength. Overall, these results demonstrate that while harmonic confinement stabilises and localises the soliton dynamics, the essential features of dark–bright soliton formation and coherent intercomponent coupling persist even in the homogeneous setting, governed primarily by the interplay of spin–orbit and Rabi interactions.

Building on the above analysis of quench-induced soliton dynamics in both trapped and trapless configurations, it is essential to develop a more systematic understanding of how the interplay between spin-orbit coupling and Rabi coupling governs the spatiotemporal evolution of the condensate. In particular, varying these coupling strengths allows one to explore the transition from weakly perturbed states to strongly modulated and propagating density structures, as well as to distinguish the role of external confinement. Figure~\ref{new:fig:9} presents a comprehensive comparison of the component densities $|\psi_{\uparrow}|^{2}$ and $|\psi_{\downarrow}|^{2}$ for different values of $k_{L}$ and $\Omega$, shown both in the presence (first and second columns) and absence (third and fourth columns) of the harmonic trap.

For the case $k_{L}=1$ and $\Omega=0$ [panels (a1–d1)], the system remains in a relatively simple dynamical regime. In the presence of the trap, the condensate exhibits confined and weakly modulated density profiles with minimal temporal variation, reflecting the absence of intercomponent coupling. In contrast, without the trap, the densities spread gradually, even though no significant internal modulation develops, indicating that spin–orbit coupling alone is insufficient to generate complex dynamics in the absence of Rabi driving.

A markedly different behavior emerges for $k_{L}=2$ and $\Omega=5$ [panels (a2–d2)]. In the trapped case, both components display enhanced density modulations and oscillatory patterns, arising from the combined effect of stronger spin–orbit and Rabi couplings. The confinement restricts large-scale motion, leading instead to pronounced internal structure formation. In the absence of the trap, however, the condensate develops propagating density waves with clear signatures of motion and spatial redistribution, highlighting the role of Rabi coupling in driving coherent intercomponent dynamics and the absence of confinement in enabling transport.

For even stronger spin-orbit coupling, $k_{L}=4$ with $\Omega=5$ [panels (a3–d3)], the system enters a highly dynamical regime. In the presence of the trap, the densities exhibit strong spatial modulations and complex oscillatory behaviour, indicating significant coupling-induced restructuring of the condensate. Without the trap, these effects are further amplified, leading to rapidly propagating and strongly modulated density patterns with pronounced asymmetry between the components. This demonstrates that increasing $k_{L}$ enhances the effective momentum transfer and nontrivial coupling between spin and motion, thereby enriching the dynamical response.

Overall, we find that the trapping potential and coupling parameters play a crucial role in determining the condensate dynamics. While the harmonic trap stabilises and localises the density evolution, preventing large-scale propagation, the absence of confinement allows for freely evolving, strongly modulated structures. At the same time, the combined increase in spin–orbit and Rabi coupling strengths drives the system from a weakly perturbed regime to one characterised by complex spatiotemporal patterns, underscoring the rich nonlinear dynamics inherent in spin–orbit coupled Bose–Einstein condensates.

\begin{figure*}[!htb]
\includegraphics[width=0.99\linewidth]{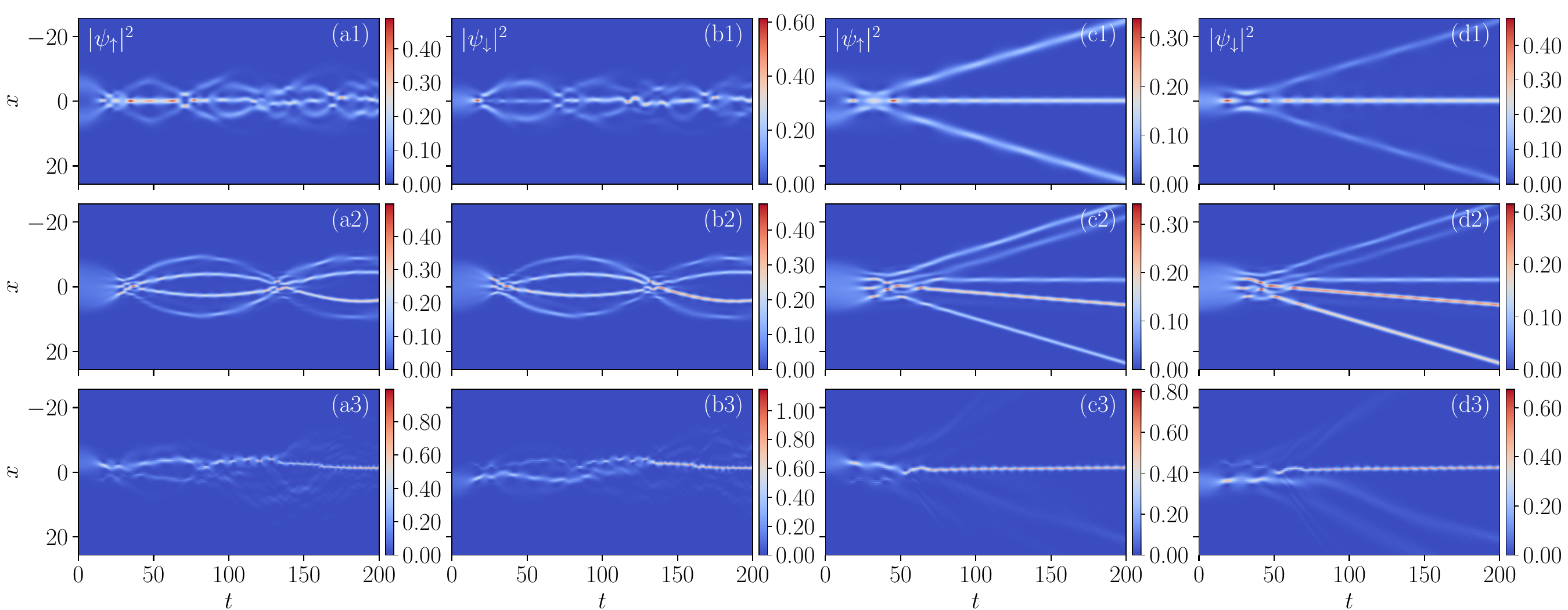}
\caption{ Spatiotemporal evolution of component densities $|\psi_{\uparrow}|^2$ and $|\psi_{\downarrow}|^2$ following an interaction quench into the immiscible regime ($g_{\uparrow\downarrow}^2 > g_{\uparrow\uparrow}g_{\downarrow\downarrow}$). Left (right) columns correspond to trapped (untrapped) dynamics. Top and bottom panels in each set show spin-up and spin-down densities, respectively. Coupling parameter sets $(k_L,\Omega)$ correspond to: (a1)--(d1) $(1,0)$, (a2)--(d2) $(2,5)$, and (a3)--(d3) $(4,5)$.}
\label{new:fig:10}
\end{figure*}

The corresponding energy dynamics (not shown) reveal that, in the presence of a trapping potential, the total energy remains conserved, with only minor redistribution occurring among the kinetic, potential, and interaction components as the coupling strengths are varied. In contrast, in the absence of the trap, the system undergoes significant expansion, which leads to a redistribution of energy between the kinetic and interaction components, reflecting the lack of a confining potential. These observations provide a baseline understanding of the equilibrium and dynamical properties of the system under repulsive interactions.

Having established the impact of SO and Rabi coupling on the dynamics of the DB soliton, we now focus on investigating the effects of a sudden quench in the interaction strength on the dynamics of the condensate.

\subsection{Repulsive-to-attractive interaction quench dynamics}

To probe the nonlinear response of the system, we consider a sudden quench from repulsive to attractive interactions. The initial ground state is prepared via imaginary-time propagation with $g_{\uparrow\uparrow} = g_{\downarrow\downarrow} = 2.0$ and $g_{\uparrow\downarrow} = g_{\downarrow\uparrow} = 2.7$. At $t=0$, the interaction parameters are abruptly switched to $g_{\uparrow\uparrow} = g_{\downarrow\downarrow} = -2.0$ and $g_{\uparrow\downarrow} = g_{\downarrow\uparrow} = -2.7$, and the system is subsequently evolved in real time.

This quench drives the condensate far away from equilibrium, resulting in rapid density redistribution and the emergence of localised structures. As shown in figure~\ref{new:fig:10}, the ensuing dynamics depend sensitively on the coupling parameters and the presence of external confinement.

For weak coupling ($k_L = 1$, $\Omega = 0$), the post-quench dynamics exhibit moderate density focusing. In the trapped system, the condensate undergoes irregular oscillations without forming stationary localised structures [figures \ref{new:fig:10}(a1) and (b1)]. In contrast, in the absence of confinement, the density contracts into a small number of localised peaks that persist over time [figures \ref{new:fig:10}(c1) and (d1)], indicating enhanced self-trapping facilitated by free expansion.

At intermediate coupling ($k_L = 2$, $\Omega = 5$), the nonlinear focusing is significantly enhanced. The density rapidly fragments into multiple localised structures resembling transient soliton trains. Under harmonic confinement, these structures undergo strong interactions characterised by repeated merging and splitting events [figures \ref{new:fig:10}(a2) and (b2)]. In free space, however, the fragments remain more spatially separated, leading to more clearly defined multi-soliton configurations [figures \ref{new:fig:10}(c2) and (d2)].

For stronger spin–orbit coupling ($k_L = 4$, $\Omega = 5$), the dynamics become more structured while remaining nonstationary. The initial density focusing is delayed, followed by the formation of fewer but more robust localised peaks. In the trapped case, these structures continue to interact and reshape due to confinement [figures \ref{new:fig:10}(a3) and (b3)], whereas in the absence of a trap, the system evolves toward a reduced number of dominant localised modes [figures \ref{new:fig:10}(c3) and (d3)].

We find that the repulsive-to-attractive interaction quench induces a wide variety of nonlinear phenomena, including density focusing, fragmentation, and soliton formation. The trapping potential primarily governs the interaction and recombination dynamics of the emergent structures, while in free space, the evolution is dominated by dispersion and spatial separation, leading to more persistent localised configurations.

\subsubsection{Dynamics of the dark-bright soliton for Symmetric interactions}
\label{sec:symmetric}

We now examine the dynamics under fully symmetric nonlinear interactions, with $g_{\uparrow\uparrow} = g_{\downarrow\downarrow}$ and $g_{\uparrow\downarrow} = g_{\downarrow\uparrow}$. Unless otherwise stated, the initial state is obtained via imaginary-time propagation with repulsive interactions, and the subsequent dynamics are studied in real time following a sudden quench of the interaction strength.

\begin{figure*}[!htb]
\includegraphics[width=\linewidth]{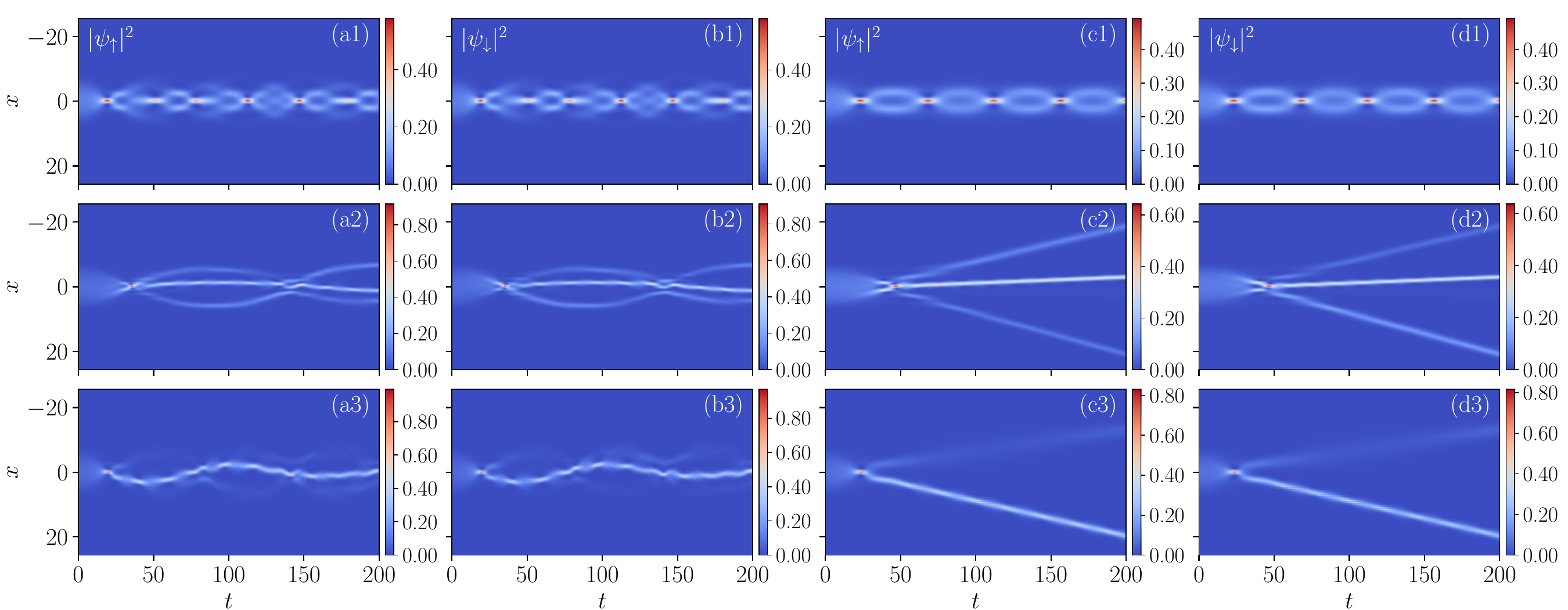}
\caption{ Spatiotemporal evolution of component densities $|\psi_{\uparrow}|^2$ and $|\psi_{\downarrow}|^2$ following an interaction quench from repulsive ($g_{ij} = 1.0$) to attractive ($g_{ij} = -1.0$) symmetric couplings. Left columns: harmonically trapped system; right columns: untrapped system. Top (bottom) panels in each set show spin-up (spin-down) densities. Coupling parameter sets $(k_L,\Omega)$ correspond to: (a1)--(d1) $(1,0)$, (a2)--(d2) $(2,5)$, and (a3)--(d3) $(4,5)$.}
\label{new:fig:11}
\end{figure*}

The primary physical mechanism driving wave-pattern formation and multi-soliton fragmentation across these interaction quenches is modulational instability (MI). When the interaction parameters $g_{ij}$ are abruptly quenched into either the immiscible regime ($g_{\uparrow\downarrow}^2 > g_{\uparrow\uparrow} g_{\downarrow\downarrow}$) or the attractive regime ($g_{ij} < 0$), background continuum modes develop complex Bogoliubov--de Gennes eigenvalues ($\text{Im}(\omega) > 0$). This MI amplifies small spatial fluctuations, causing the background condensate and uniform dark--bright soliton tails to break up into high-frequency spatial modulations, stripe patterns, and periodic soliton trains, as detailed below for figures~\ref{new:fig:11}, \ref{Symmetric-main/Kl4Om5-M1B}, and \ref{Symmetric-main/Kl4Om5-M1AG}.


For symmetric nonlinearities, we find that different initial configurations yield qualitatively identical density distributions, differing only in their corresponding energies. Consequently, we restrict our discussion to the emergent dynamical features following a quench from repulsive to attractive interactions, which captures the essential physics of pattern formation and nonlinear evolution.

In the present subsection, we focus on the case where both intra- and interspecies interactions are attractive, i.e., $g_{\uparrow\uparrow} = g_{\downarrow\downarrow} = g_{\uparrow\downarrow} = g_{\downarrow\uparrow} = -1.0$, and analyse the resulting dynamics for different coupling parameters. The corresponding spatiotemporal evolution of the component densities is summarised in figure \ref{new:fig:11}, which presents results both in the presence and absence of the harmonic trap.

\subsubsection{Attractive intra- and interspecies interactions}

Figure~\ref{new:fig:11} shows the time evolution of the densities $|\psi_{\uparrow}|^2$ and $|\psi_{\downarrow}|^2$ for three representative parameter sets: $(k_L,\Omega) = (1,0)$, $(2,5)$, and $(4,5)$. The left two columns correspond to the trapped system, while the right two columns represent the trap-free case. In each case, the upper (lower) panel within a column shows the evolution of the spin-up (spin-down) component.

For $(k_L,\Omega) = (1,0)$ [figure \ref{new:fig:11}(a1)–(b1) and (c1)–(d1)], the system exhibits a robust breather-like soliton in both trapped and untrapped configurations. The density undergoes periodic oscillations without significant dispersion, indicating a stable non-linear bound state that is largely insensitive to the presence of the trap.

For $(k_L,\Omega) = (2,5)$ [figure \ref{new:fig:11}(a2)–(b2) and (c2)–(d2)], the dynamics become considerably richer. A rapid localisation occurs at intermediate times, leading to the formation of a multi-soliton structure. In the trapped case, this manifests as a three-soliton fragmentation, where a dominant central soliton is accompanied by two weaker side peaks. These structures undergo repeated merging and splitting, resulting in a non-periodic beating pattern. In the absence of the trap, a qualitatively similar fragmentation is observed, although the localisation occurs slightly later in time and the resulting soliton train exhibits reduced confinement-induced recombination.

For $(k_L,\Omega) = (4,5)$ [figure \ref{new:fig:11}(a3)–(b3) and (c3)–(d3)], the dynamics are dominated by the formation of stripe-like solitonic structures. In the trapped system, these appear as oscillating stripe solitons, whereas in the untrapped case, a single high-density stripe soliton emerges and propagates with minimal deformation. This behaviour is consistent with earlier observations of stripe solitons in spin--orbit-coupled condensates~\cite{Gangwar2022}. We have shown the importance of coupling parameters $(k_L,\Omega)$ in governing the transition from simple breather dynamics to complex multi-soliton fragmentation and stripe formation. While the harmonic trap modifies the quantitative features of the evolution, particularly the recombination and confinement of solitons, the qualitative dynamical regimes remain robust across both trapped and homogeneous settings.

\begin{figure}[!htb]
\centering\includegraphics[width=0.99\linewidth]{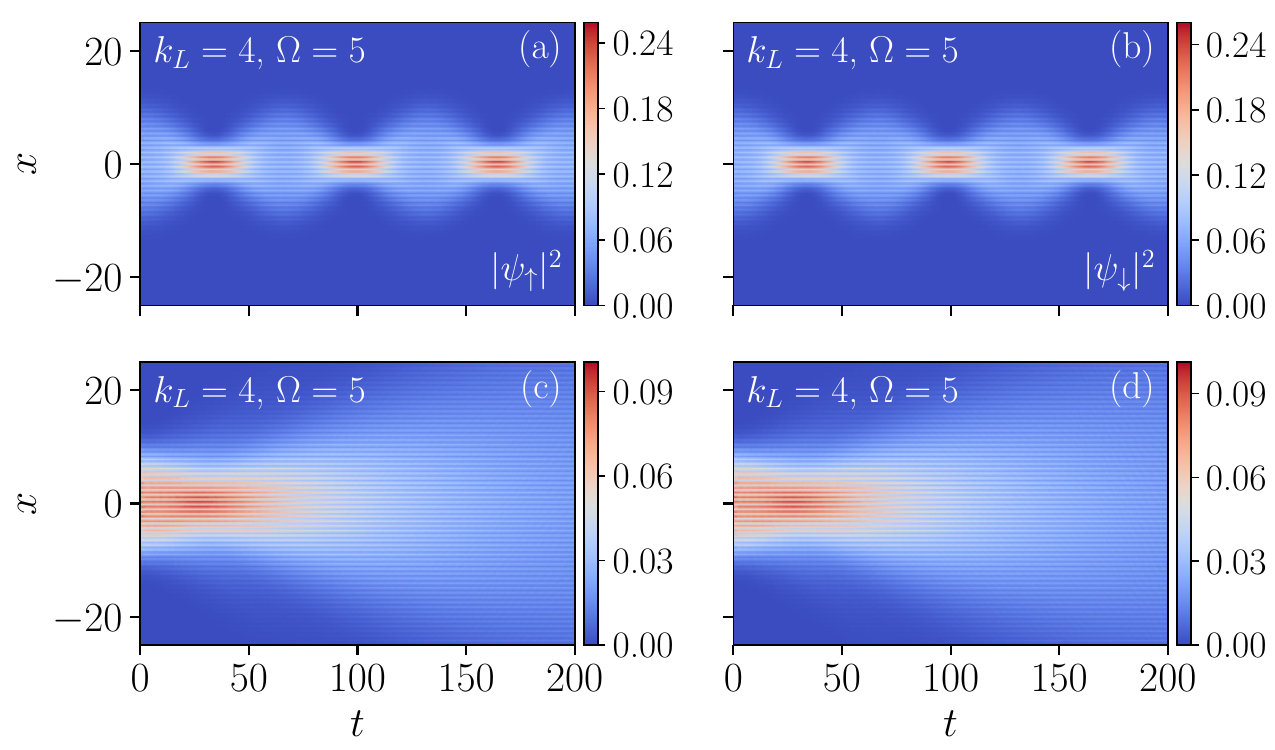}
\caption{ Spatiotemporal evolution of component densities $|\psi_{\uparrow}|^2$ and $|\psi_{\downarrow}|^2$ for $(k_L,\Omega) = (4,5)$ following an interaction quench to mixed nonlinearities ($g_{\uparrow\uparrow} = g_{\downarrow\downarrow} = 1.0$, $g_{\uparrow\downarrow} = g_{\downarrow\uparrow} = -1.0$). Panels (a,b) show trapped dynamics; panels (c,d) show untrapped dynamics. Initial ground state prepared with $g_{ij} = 1.0$.}
\label{Symmetric-main/Kl4Om5-M1B}
\end{figure}

Next we consider mixed nonlinearities with $ g_{\uparrow\uparrow} = 1.0 $, $ g_{\uparrow\downarrow} = g_{\downarrow\uparrow} = -1.0 $, and $ g_{\downarrow\downarrow} = 1.0 $, while keeping the coupling strengths fixed at $ k_{L}=4 $ and $ \Omega=5 $. The initial state is prepared as the ground state corresponding to purely repulsive interactions, and the subsequent dynamics are induced by a sudden quench to the mixed interaction regime.

Figure~\ref{Symmetric-main/Kl4Om5-M1B} presents the resulting time evolution of the component densities. In the presence of the harmonic trap [panels (a) and (b)], the densities $ |\psi_{\uparrow}|^2 $ and $ |\psi_{\downarrow}|^2 $ exhibit well-defined breathing stripe soliton patterns. These structures are characterised by periodic modulation in both space and time, arising from the combined effects of spin–orbit coupling, Rabi-induced intercomponent coherence, and attractive interspecies interaction. The harmonic confinement plays a stabilising role by restricting spatial expansion, thereby sustaining coherent and long-lived oscillatory dynamics.

In contrast, in the absence of the trap [panels (c) and (d)], the system undergoes significant expansion, and the dynamics are dominated by self-interference effects. The densities develop intricate interference fringes as the wave packets expand and overlap during evolution. Without external confinement, the kinetic energy drives the spreading of the condensate, while the underlying coupling mechanisms imprint phase coherence that manifests as structured interference patterns.

We find that while attractive interspecies interactions favour the formation of bound and modulated solitonic structures, the presence of a trapping potential is essential for their stabilisation. In its absence, the system transitions to a regime dominated by expansion and self-interference, highlighting the rich interplay between interactions, coupling, and confinement in governing the emergent nonlinear dynamics, as also reported for bright soliton dynamics in SO coupled BEC~\cite{Ravisankar2025, Gangwar2023}.

\begin{figure}[!htb]
\centering\includegraphics[width=0.99\linewidth]{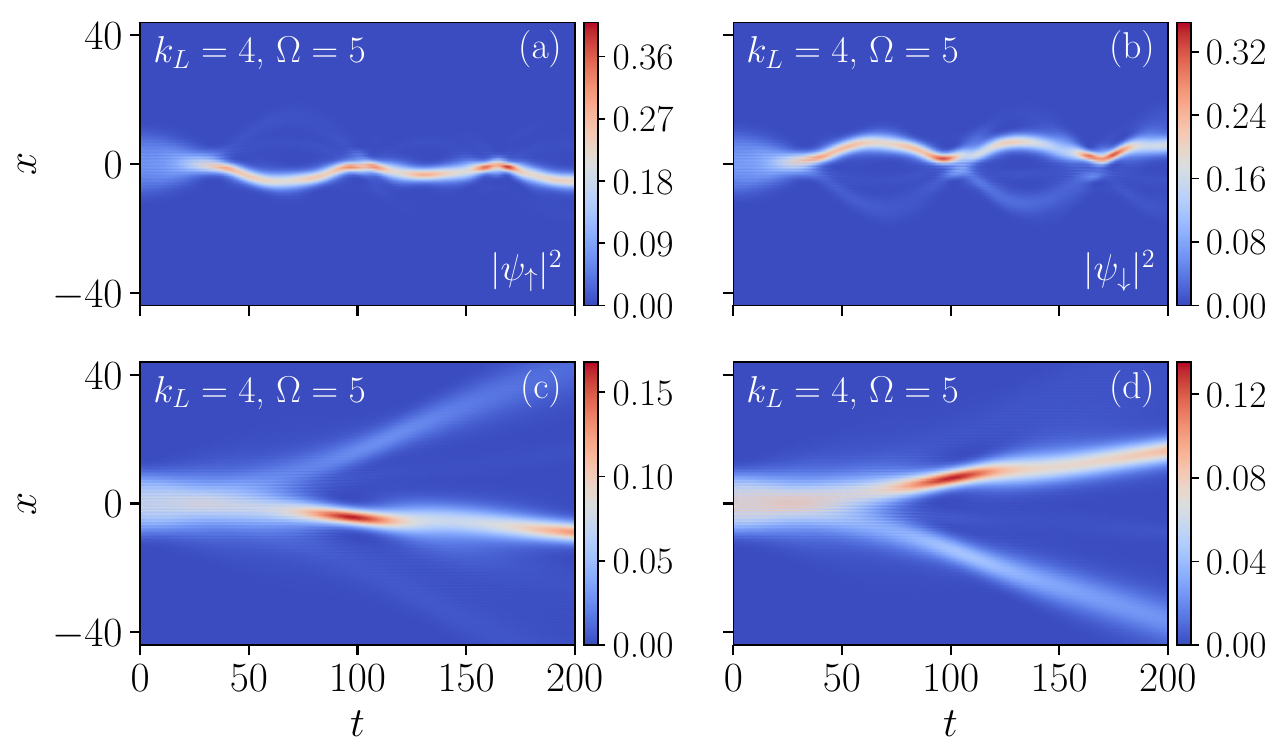}
\caption{ Spatiotemporal evolution of component densities $|\psi_{\uparrow}|^2$ and $|\psi_{\downarrow}|^2$ for $(k_L,\Omega) = (4,5)$ following an interaction quench to sign-reversed nonlinearities ($g_{\uparrow\uparrow} = g_{\downarrow\downarrow} = -1.0$, $g_{\uparrow\downarrow} = g_{\downarrow\uparrow} = 1.0$). Panels (a,b) show trapped dynamics; panels (c,d) show untrapped dynamics. Initial ground state prepared with $g_{ij} = 1.0$.}
\label{Symmetric-main/Kl4Om5-M1AG}
\end{figure}

Continuing from the above discussion, we now focus on the specific dynamical evolution of dark–bright solitons under a quench to mixed nonlinearities at fixed coupling parameters $k_L = 4$ and $\Omega = 5$. Figure~\ref{Symmetric-main/Kl4Om5-M1AG} illustrates the corresponding time evolution of the component densities. Panels (a) and (b) show the evolution of $|\psi_{\uparrow}|^2$ and $|\psi_{\downarrow}|^2$, respectively, in the presence of the harmonic trap. In this case, the initial ground state is prepared with purely repulsive nonlinearities $g_{\uparrow\uparrow} = g_{\uparrow\downarrow} = g_{\downarrow\uparrow} = g_{\downarrow\downarrow} = 1.0$, and the subsequent real-time dynamics are generated by switching to mixed nonlinearities $g_{\uparrow\uparrow} = -1.0$, $g_{\uparrow\downarrow} = g_{\downarrow\uparrow} = 1.0$, and $g_{\downarrow\downarrow} = -1.0$. Under harmonic confinement, the system supports robust dark–bright soliton structures that undergo breathing-type oscillations while remaining spatially localised. The interplay between spin–orbit coupling, Rabi coupling, and the sign-changing nonlinear interaction leads to periodic modulation of both components, yet the trap ensures long-term coherence and prevents dispersive spreading.

In contrast, panels (c) and (d) present the corresponding dynamics in the absence of the harmonic trap under identical interaction quench conditions. Here, the confinement-free evolution lets the solitonic wave packets expand freely, leading to enhanced spatial spreading and the emergence of pronounced self-interference patterns. While the dark–bright soliton character remains identifiable at intermediate times, the lack of external confinement gradually weakens localisation, resulting in a transition toward a more dispersive regime. The asymmetric nonlinear interaction further amplifies this effect by inducing differential evolution in the two spin components, thereby modifying their relative phase and density profiles. Interestingly, we find that the intrinsic coupling mechanisms are sufficient to generate coherent nonlinear structures; the harmonic trap is essential for sustaining their localisation and preventing long-time dispersive decay in the presence of interaction sign reversal.

\section{Summary and conclusion}
\label{sec:6}

In this paper, we have presented a detailed numerical and analytical investigation of the ground-state properties and real-time dynamics of dark--bright solitons in a binary Bose--Einstein condensate with synthetic spin--orbit and Rabi couplings. The behaviour of these nonlinear excitations is shaped by the complex interplay among interaction strengths, synthetic gauge fields, and external confinement.

In the absence of synthetic couplings, the trapped system exhibits the expected miscible--immiscible transition as the interspecies interaction exceeds a critical threshold. Within this regime, dark--bright solitons display robust oscillatory motion, consistent with their effective particle-like dynamics. The inclusion of spin--orbit coupling qualitatively alters the evolution: under harmonic confinement, it induces persistent internal density oscillations whose amplitude increases with coupling strength, while in free space, it drives counter-propagating motion of the spin components, reflecting spin--momentum locking.

The addition of Rabi coupling introduces coherent interconversion between components, leading to breather-like dynamics in trapped systems with frequencies determined by the coupling strengths. In the absence of confinement, the combined action of spin--orbit and Rabi couplings produces oscillatory expansion dynamics, accompanied by sustained oscillations in the interaction energy, while the total energy remains conserved within numerical accuracy. {Furthermore, by evaluating the continuity relations for mass and spin current densities, we have mapped the internal spin dynamics onto an internal Josephson junction framework, where the synthetic gauge field acts as a continuous spatial momentum bias driving asymmetric spin-current pulses across the soliton notch.

When synthetic couplings are included at the level of ground-state preparation, the system supports distinct phases, namely the plane-wave phase ($k_L^2 < \Omega$) and the stripe phase ($k_L^2 > \Omega$). Bogoliubov--de Gennes analysis reveals that trapped systems may exhibit dynamical instabilities at intermediate coupling strengths, whereas uniform systems remain stable over the explored parameter regime. Subsequent real-time evolution from these dressed states leads to modulated plane-wave or stripe patterns in confined geometries, and to expanding or localised structures in free space.

The richest nonlinear dynamics emerge following interaction quenches. We have identified modulational instability as the core physical mechanism driving the break-up of background continuum modes and dark--bright soliton tails. Transitions from repulsive to attractive interactions generate {modulational-instability-induced} multi-soliton fragmentation and irregular, non-periodic oscillations. Mixed-sign interaction regimes give rise to qualitatively distinct behaviours, including breathing stripe patterns under confinement and expanding self-interference structures in free space, as well as fragmented or bifurcating density profiles depending on the specific interaction configuration.

Overall, our results demonstrate that dark--bright soliton dynamics in spin--orbit--coupled binary condensates can be effectively controlled through the combined tuning of interactions, synthetic couplings, and external trapping. While confinement favours oscillatory and breather-like behaviour, free-space evolution enables directional separation and expansion. Spin--orbit and Rabi couplings introduce internal spin dynamics and stabilise phases with characteristic density modulations, whereas interaction quenches provide access to strongly nonequilibrium regimes featuring pattern formation and fragmentation. The agreement between Bogoliubov analysis, {current density formulations,} and nonlinear time evolution establishes a consistent framework for understanding stability and dynamical responses in these multicomponent quantum fluids.

\acknowledgements
KR acknowledges financial support from UGC-SJSGC. 
RR wishes to acknowledge the financial assistance from DST-CURIE (DST-CURIE-PG/2022/54) and ANRF (DST-CRG/2023/008153).
P.M. acknowledges the financial support from MoE RUSA 2.0 (Bharathidasan University -- Physical Sciences). 



\providecommand{\newblock}{}

\end{document}